\documentstyle[epsf]{mn}
\input epsf.sty
\def\vmed{$\langle v \rangle$~}
\def\hmpc{~h$^{-1}$ Mpc~}

\def\hmendue{~h$^{2}$ Mpc$^{-2}$~}

\def\hmagm{$-5\log h$~}
\def\parn{\par\noindent}
\title[A study of the core of the Shapley Concentration: III]
 { A study of the core of the Shapley Concentration: \\
 III. Properties of the clusters in the A3558 complex 
\thanks{based on observations collected at the European Southern
        Observatory, La Silla, Chile.} } 
\author[S. Bardelli et al.]{
S. Bardelli$^{1}$,
E. Zucca$^{2,3}$,
G. Zamorani$^{2,3}$,
G. Vettolani$^{3}$
\&
R. Scaramella$^{4}$
\\ $^1$ Osservatorio Astronomico di Trieste, 
via Tiepolo 11, I--34131 Trieste, Italy
\\ $^2$ Osservatorio Astronomico di Bologna, 
via Zamboni 33, I--40126 Bologna, Italy
\\ $^3$ Istituto di Radioastronomia del CNR, 
via Gobetti 101, I--40129 Bologna, Italy
\\ $^4$ Osservatorio Astronomico di Roma, 
via Osservatorio 2, I--00040 Monteporzio Catone (RM), Italy
\\ E-mail: bardelli@astrts.oat.ts.astro.it
}
\date{Received 00 - 00 - 0000; accepted 00 - 00 - 0000}
\begin{document}
\maketitle

\begin{abstract}
The Shapley Concentration is the richest supercluster of clusters in the nearby 
universe and its core is a remarkable complex formed by the ACO clusters 
A3558, A3562 and A3556, and by the two minor groups SC 1327-312 and SC 1329-314.
This structure has been studied in various wavelength bands, revealing that 
it is probably dynamically very active.
In this paper we present 174 new galaxy redshifts in this cluster complex,
which are added to the sample of 540 already existing velocities.
 The large number of redshifts permits a more accurate and robust 
analysis of the dynamical parameters of the clusters. In particular, we discuss
the complex velocity distribution of A3558, the bimodal distribution of 
A3556 and SC 1329-313, and calculate the mean velocity and velocity dispersion 
of A3562. Moreover, for the three ACO clusters we derive the luminosity 
functions adopting a new fitting technique, which takes into account 
the galaxy density profiles. 
\end{abstract}

%
%
%-------------------------------------------------------------------------------
%
\section{Introduction}
The Shapley Concentration supercluster, at a distance of $\sim 140$\hmpc 
(h=H$_0/100$), is a very interesting region, which makes possible the 
detailed study of the formation and evolution of clusters of galaxies and of 
the role played by the environment on these phenomena.
The high local density contrast, as traced by the distribution of
Abell--ACO clusters (Scaramella et al. 1989; Zucca et al. 1993), X--ray
clusters (Lahav et al. 1989) and both optical (Raychaudhury et al. 1991)
and IRAS galaxies (Allen et al. 1990), suggests the presence of
high peculiar velocities of the order of $\sim 1000$ km/s (Branchini \& Plionis 
1996). 
This fact, added to the richness in clusters 
of this supercluster (it is the richest one within $300$\hmpc, 
Zucca et al. 1993), leads to an increase of the 
cross--section for interactions, like cluster--cluster and group--cluster
mergings. Raychaudhury et al. (1991), noting that the percentage of
multiple nuclei clusters is much higher than in the field, suggested a merging 
rate between 1.5 and 3 times higher than elsewhere. 
\par
The core of the Shapley Concentration is individuated by a remarkable chain
formed by the three ACO (Abell et al. 1989) clusters A3556, A3558 and A3562, 
already noted by
Shapley (1930). This complex is an aligned structure, elongated  $\sim 3^o$
in the East--West direction, with a comoving size of $\sim 7.5$\hmpc.
As noted by Metcalfe et al. (1994), A3556 (richness class 0) and A3562
(richness class 2) lie within the Hubble
turnaround radius of A3558, and probably they are going to merge with the 
latter. 
\par
The bi--dimensional distribution of galaxies suggests that these clusters
are strongly interacting and are forming a single connected structure (Bardelli 
et al. 1994, hereafter Paper I; Metcalfe et al. 1994).
In particular, it is possible to individuate a number of secondary subclumps
surrounding A3558 (see Figure 14 of Paper I), revealing the complex 
dynamical state of this structure.
\par
In Paper I we presented the results of an extensive spectroscopic survey 
in this area, using a sample of $\sim 300$ new redshifts, to be added
to the $200$ from literature (Metcalfe et al. 1987; Teague et al. 1990).
We confirmed that the three ACO clusters are at the same distance and
are aligned in a filament approximately perpendicular to the line of sight. 
Moreover
we studied two subcondensations between A3558 and A3562, near the position
of the poor cluster SC 1329--314; the reality of these groups was
confirmed by observations in the X--ray band by Bardelli et al. (1996, 
hereafter Paper II) and Breen et al. (1994), who detected extended emission
from the hot gas trapped in their gravitational wells.
A significant X--ray emission connecting A3558 with the nearest group 
(dubbed SC 1327--312 by Breen et al.) has been detected (see Paper II), 
confirming the presence of strong gravitational interactions between these 
components. 
This fact reinforces the suggestion by Metcalfe et al. (1994) that 
there was already an encounter between A3562, A3558 and the two groups.
These collisions, expected in a hierarchical scenario of formation of cosmic 
structures, have been studied through numerical simulations by 
McGlynn \& Fabian (1984) and Roettiger et al. (1993) and have probably  been
observed in A2256 (Briel et al. 1991) and Coma (Burns et al. 1994).
\par
Moreover, the merging between a cluster and a group could explain the formation 
of the central dominant galaxies. Indeed, following Tremaine (1990), such 
galaxies are likely to form in low velocity dispersion systems (such as groups),
which are then captured by clusters and deposited at their centers through 
dynamical friction.
\par
For these reasons the detailed study of interacting clusters, and in particular 
of their dynamical state, is important: this can be done with the
simultaneous use of bi--dimensional and velocity information. 
Therefore it is necessary to cover wide areas with multifiber spectroscopy, 
in order to obtain large redshift samples, which are necessary to study all
the details of the galaxy distribution.
\par
In this paper we present various properties (dynamical parameters, luminosity
functions, density profiles) of the clusters in the A3558 complex, using also
$\sim 200$ new galaxy redshifts. 
The paper is organized as follows: in Sect.2 we present the galaxy sample
and in Sect.3 we describe the analyses applied to the clusters; the results 
for each cluster are given in Sect.4 and finally in Sect.5 there is the
summary.
%-------------------------------------------------------------------------------
% FIGURE 1. 
\begin{figure}
\epsfysize=8.5cm
\epsfbox{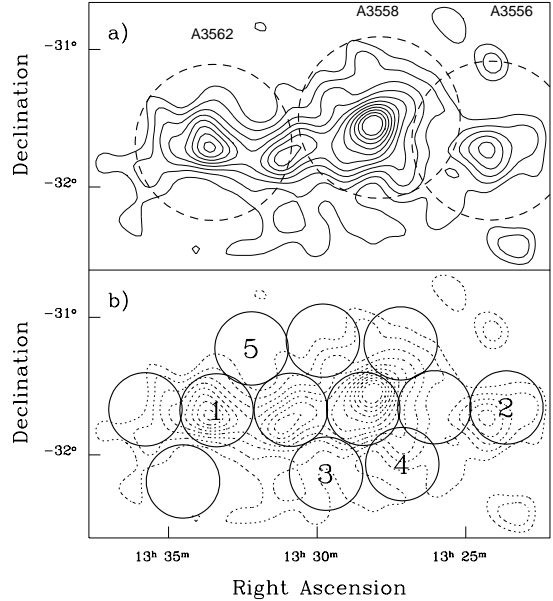}
\caption[]{
a) Isodensity contours of the core of the Shapley Concentration
in an area of $\sim 3^o.2 \times 1^o.4$. 
The figure refers to galaxies with $b_J \le 19.5$ and binned in 
$2$  $\times 2$ arcmin$^2$ bins; the data have been smoothed with a Gaussian 
with a FWHM of $6$ arcmin. For the three Abell clusters circles
of one Abell radius have been drawn (dashed curves); the poor cluster
SC 1329--314 is the peak between the clusters A3558 and A3562. 
\parn
b) The same as in panel a), with superimposed the observed OPTOPUS 
fields: circles with numbers refer to the new observations, the others 
correspond to the sample of Paper I. }
\label{fig:isodense}
\end{figure}
%-------------------------------------------------------------------------------

%-------------------------------------------------------------------------------
%
\section{The sample}

%-------------------------------------------------------------------------------
%
\subsection{The galaxy catalogue}

Figure \ref{fig:isodense}a shows the isodensity contours for the objects with 
$b_J\le 19.5$
from the COSMOS/UKST galaxy catalogue (Yentis et al. 1992), in a region of 
$3^o.2\times 1^o.4$, corresponding to $13^h 22^m 06^s<\alpha(2000) < 13^h 37^m 
15^s$ and $-32^o 22' 40''<\delta(2000) < -30^o 59' 30''$. 
This region is part of the UKSTJ plate 444 and contains 2241
galaxies within the chosen limits. The radius of the dashed circles, centered
on A3562, A3558 and A3556, corresponds to 1 Abell radius ($\sim 1.5$ \hmpc); 
the overdensity between A3562 and A3558 corresponds to SC 1329--314. Figure 
\ref{fig:isodense}b shows, superimposed on the galaxy contours, the position of 
the fields observed with the multifiber spectrograph OPTOPUS (solid circles with
32 arcmin diameter):
the five numbered fields correspond to the new observations, while the others
refer to the data reported in Paper I.
Field 1, which corresponds to the center of A3562, was observed twice in order to compensate
the completely failed previous observations, and field 2 is a further 
observation of the center of A3556, previously observed as field 9 (see Paper I). 
\par
The coordinates and the observation dates of the new fields are reported in
Table 1.

%-------------------------------------------------------------------------------
% TABLE 1.
\begin{table}
\caption[]{ Centers of the observed fields }
\begin{flushleft}
\begin{tabular}{llll}
\hline\noalign{\smallskip}
Field & $\alpha$ (2000) & $\delta$ (2000) & Observation date \\
\noalign{\smallskip}
\hline\noalign{\smallskip}
  1  & 13 33 30 & -31 40 00 & 21 Feb 1993 \\
  2  & 13 23 36 & -31 39 21 & 21 Feb 1993 \\
  3  & 13 29 45 & -32 07 43 & 22 Feb 1993 \\
  4  & 13 27 15 & -32 05 00 & 24 Feb 1993 \\
  5  & 13 32 15 & -31 12 17 & 23 Feb 1993 \\
\noalign{\smallskip}
\hline
\end{tabular}
\end{flushleft}
\end{table}
%-------------------------------------------------------------------------------

%-------------------------------------------------------------------------------
%
\subsection{Observations and data reduction}

The spectroscopic observations were performed at the 3.6m ESO telescope 
at La Silla, equipped with the OPTOPUS multifiber spectrograph (Lund 1986), on 
the nights of 21--22--23--24 February 1993.
The OPTOPUS multifiber spectrograph uses bundles of 50 optical 
fibers, which can be set within the field of the Cassegrain focal plane 
of the telescope; this field has a diameter of $32$ arcmin, and each fiber
has a projected size on the sky of $\sim 2.5$ arcsec.
We used the ESO grating $\#\ 15$ 
($300$ lines/mm and blaze angle of $4^o 18'$) allowing a dispersion of 
$174$ \AA /mm (corresponding to a resolution of $\sim 12$ \AA) in the 
wavelength range from $3700$ to $6024$ \AA. 
The detector was the Tektronic $512\times 512$ CB CCD (ESO $\# 32$) with a 
pixel size of $27\ \mu$m, corresponding to $4.5$ \AA, i.e. a velocity bin of 
$\sim 270$ km/s at $5000$ \AA. 
This detector has a good responsive quantum function in the blue ($\sim 70\%$
at $4000$ \AA), where there are the calcium and [OII] lines.
Following Wyse \& Gilmore (1992), we dedicated 5 fibers to sky measurements, 
remaining with 45 fibers available for the objects.
The observing time for each field was one hour, split into two half--hour
exposures in order to minimize the effects due to the ``cosmic'' hits. The 
observing sequence was: a 30 seconds exposure of a quartz--halogen white lamp,
a 180 seconds exposure of the Phillips Helium and Neon arcs, then the first and 
the second field exposures, and again the arcs and the white lamp. 
\par
The reduction steps are described in Paper I. However, it could be important to
stress that we normalized the fiber transmission dividing each spectrum by
the continuum--subtracted flux of the sky emission line [OI]$\lambda$ 5577. 
This procedure assumes that the sky emission does not change significantly
on angular scales of the order of $\sim 30$ arcmin (Wyse \& Gilmore 1992). 
\par 
We obtained estimates of the radial velocity of galaxies using the 
cross--correlation method as implemented in the IRAF 
\footnote{IRAF is distributed by 
KPNO, NOAO, operated by the AURA, Inc., for the National Science Foundation.}
task RVSAO (XCSAO, Kurtz et al. 1992). 
The galaxy spectra were cross--correlated with those of $8$ stellar templates
observed with the same instrumental set up. 
The adopted velocity for each galaxy is the value given by the template
which gives the minimum cross--correlation error. For spectra with strong
emission lines we measured an ``emission velocity" using 
the EMSAO program in the same IRAF task RVSAO.

%-------------------------------------------------------------------------------
%
\subsection{The new redshift data}

From the total number of spectra ($266$), it was possible to obtain $214$
velocity estimates: $40$ objects turned out to be stars ($\sim 19\%$ of the
reliable spectra), leaving us with $174$ new galaxy redshifts. The high number 
of failed spectra ($\sim 19\%$ of the total number of spectra) is mainly due to
technical problems in the second exposure (fainter galaxies) of field $\# 1$:
if we do not consider this exposure, the total number of spectra is 222 with
19 failed spectra ($\sim 9\%$).

In Table 2 we list the objects with velocity determination. Column (1) is the 
sequential number, columns (2), (3) and (4) give the right ascension (2000), 
the declination (2000) and the $b_J$ apparent magnitude, respectively, as 
reported in the COSMOS catalogue. 
Columns (5) and (6) give the heliocentric velocity ($v=cz$) 
and the internal error (in km/s); the word ``star" in column (5)
indicates those objects which turned out to be stars after the spectral
analysis. The external error can be derived by multiplying the error in
column (6) by a factor of the order of $1.6-1.9$: the lower value is
obtained by comparing repeated observations of the same galaxies
(Malumuth et al. 1992) and the higher one takes into account also the 
different reduction techniques (see the discussion in 
Paper I). 
Finally, the code ``emiss" in column (7) denotes the velocities determined from 
emission lines. 

The galaxies whose spectrum presents detectable emission lines 
(mainly [OII]$\lambda$ 3727, H$\beta$ $\lambda$ 4860, [OIII]$\lambda$ 4959 and
[OIII]$\lambda$ 5007) are $30$, corresponding 
to a percentage of $17\%$, in agreement with the average value of $16\%$ 
reported by Biviano et al. (1997a, 1997b) for the cluster galaxy population. 

Adding also the previous data, the total spectroscopic sample consists of 714 
galaxy redshifts: 511 are those used in Paper I, 29 are from Quintana et al. 
(1995) and 174 are our new determinations. 
The completeness of the spectroscopic survey at $b_J \le 19.5$ is $\sim 31\%$ 
($\sim 50\%$ at $b_J \le 18.0$) in the entire area, but it strongly varies 
with the position:
within one Abell radius from the cluster centers we have $\sim 45\%$ of the 
redshifts for A3558 ($\sim 82\%$ at $b_J \le 18.0$), $\sim 26\%$ ($\sim 55\%$) 
for A3562 and $\sim 30\%$ ($\sim 68\%$) for A3556.

\begin{figure}
\epsfysize=9cm
\epsfbox{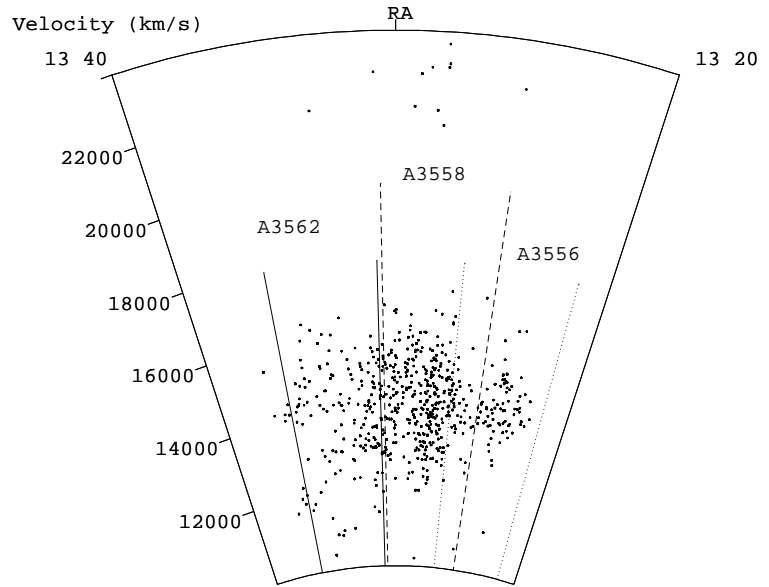}
\caption[]{ Wedge diagram of the sample of galaxies in the velocity range 
$10000 - 24000$ km/s. The coordinate range is 
$13^h 22^m 06^s < \alpha (2000)< 13^h 37^m 15^s$ and 
$-32^o 22' 40''< \delta(2000) < -30^o 59' 30''$. The three pairs of straight 
lines (solid, dashed and dotted) show the projection in right ascension of
1 Abell radius for the three clusters A3562, A3558 and A3556, respectively. }
\label{fig:wedge}
\end{figure}

In Figure \ref{fig:wedge} we show the wedge diagram of this sample in the 
velocity range $[10000-24000]$ km/s. 
This wedge, with $\sim 40\%$ more redshifts than those presented in Paper I,
confirms the connection between the clusters: the A3558 complex is a single 
structure of galaxies, ranging from A3556 to A3562.
In particular, the data now available for A3562 permits to study also 
the eastern part of the complex, where the galaxy distribution appears
to be less concentrated and the existence of a ``finger-of-God" effect 
 is not clear.

%-------------------------------------------------------------------------------
% FIGURE 2. 
%-------------------------------------------------------------------------------

%-------------------------------------------------------------------------------
% FIGURE 3. 
\begin{figure}
\epsfysize=8.5cm
\epsfbox{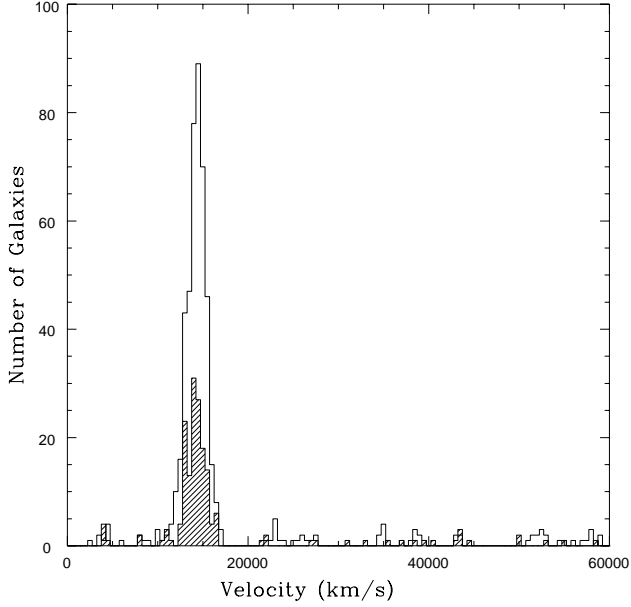}
\caption[]{Velocity distribution for the 174 new redshifts (shadowed histogram),
superimposed to the data used in Paper I (open histogram). 
It is clear the presence of the overdensity corresponding to the core of the 
Shapley Concentration (at $\sim 15000$ km/s) and a void at $\sim 20000$ km/s, 
well represented in both samples. }
\label{fig:histogram}
\end{figure}
%-------------------------------------------------------------------------------
\par
In Figure \ref{fig:histogram} the velocity distribution of the 174 new redshifts
(shadowed histogram) is shown, superimposed to the data used in Paper I.
It is clear that both samples well represent the overdensity at the
distance of the core of the Shapley Concentration ($\sim 15000$ km/s) and
the void (at $\sim 20000$ km/s), already noted in Paper I, just behind the 
structure.

%-------------------------------------------------------------------------------
%
\section{Analysis of the clusters}

%-------------------------------------------------------------------------------
\subsection{Dynamical parameters}

In order to estimate the dynamical parameters of the cluster
velocity distribution, i.e. the mean velocity \vmed and the velocity dispersion
$\sigma$, we used the biweight location and scale estimators, respectively 
(see Beers et al. 1990 for their definitions). The advantage of these 
estimators, with respect to the standard mean and dispersion, is that of
minimizing biases from interlopers, giving less weight to data with higher
distance from the median.
The derived velocity dispersions are then corrected for the broadening due to 
velocity errors, following Danese et al. (1980). 
The confidence intervals of the two estimators 
are calculated bootstrapping the data with 100 random catalogs. 

In order to find the velocity range in which the cluster members lie, we have 
assumed that the velocity distribution of cluster galaxies is Gaussian, as 
expected when the system has undergone a violent relaxation (see details in 
Paper I). For the cases in which the presence of a substructure was suspected
on the basis of either the shape estimators $a$, $b_1$, $b_2$ and $I$ (see
Bird \& Beers 1993) or a visual inspection of the velocity histogram, we 
checked if the distribution is consistent with a single Gaussian
or it is bimodal applying the KMM test (Ashman et al. 1994), using the
program kindly provided by the authors. This test gives the likelihood ratio
between the hypothesis that the dataset is better described by the sum
of two Gaussians and the null hypothesis that the dataset is
better described by a single Gaussian.
 The value of the likelihood ratio, as computed
by the program, is a rigorous estimate of the confidence limit
of the null hypothesis only when testing the existence of two Gaussians
with equal dispersions (homoscedastic distributions), but can be
taken as a reasonable approximation also in the case of different dispersions 
(heteroscedastic) (Ashman et al. 1994). However, in the cases in which
we found bimodality in our data, the bimodal distribution was significant both
in the homoscedastic and in the heteroscedastic case (if not otherwise
specified).

In the case of bimodality, the KMM program gives for each galaxy the ``a 
posteriori probability" of the group membership: after having divided the two 
groups on this basis, we have repeated the estimate of mean velocity and 
velocity dispersion for each of them following the previous criteria.

The results of the dynamical analysis of the considered clusters are reported in
Table 3. The first column lists the cluster name, the second and third columns
give the mean velocity and velocity dispersion, the fourth column reports the
number of galaxies used for the determination of the parameters and in the
fifth column there are the characteristics of the used samples. The discussion
of the results for each cluster is reported in Sect.4.

%-------------------------------------------------------------------------------
%%%%%%%%%%%%%%%%%%%%%%%%%%%%%%%%%%%%%%%
%TAB. 3 RIASSUNTO DINAMICA DEI CLUSTERS
%%%%%%%%%%%%%%%%%%%%%%%%%%%%%%%%%%%%%%%
\setcounter{table}{2} 
\begin{table}
\caption[]{ Dynamical parameters of the clusters}
\begin{flushleft}
\begin{tabular}{lllll}
\hline\noalign{\smallskip}
Name & \vmed  & $\sigma$ & $N_{gal}$ & notes \\
     & (km/s) & (km/s)   &           &       \\
\noalign{\smallskip}
\hline\noalign{\smallskip}
A3556 & 14357$^{+76}_{-76}$ & 643$^{+53}_{-43}$ & 79 & r$<$22 arcmin     \\
      & 14130$^{+42}_{-74}$ & 411$^{+76}_{-29}$ & 56 & substr. 1       \\ 
      & 15066$^{+58}_{-43}$ & 222$^{+59}_{-35}$ & 23 & substr. 2        \\ 
A3558 & 14403$^{+60}_{-55}$ & 996$^{+40}_{-36}$ &307 & r$<$1 r$_{Abell}$  \\ 
      & 14262$^{+75}_{-82}$ & 992$^{+85}_{-60}$ &155 & r$<$0.5 r$_{Abell}$ \\ 
      & 14260$^{+100}_{-67}$ & 946$^{+70}_{-80}$ &117 & $b_J>$17         \\ 
      & 14309$^{+285}_{-219}$ &1125$^{+139}_{-79}$ &31 & $b_J<$17        \\ 
      & 12993$^{+129}_{-119}$ & 310$^{+119}_{-92}$ &11 & $b_J<$17 substr. 1\\ 
      & 14871$^{+144}_{-145}$ & 455$^{+51}_{-55}$ &20 & $b_J<$17 substr. 2\\ 
SC 1329--313 & 14790$^{+114}_{-67}$ & 377$^{+93}_{-82}$ & 16 &  substr. 1 \\
            &  13348$^{+69}_{-83}$ & 276$^{+70}_{-61}$ & 21 &  substr. 2 \\
SC 1327--312 & 14844$^{+105}_{-211}$ & 691$^{+158}_{-146}$ & 24 &  \\
A3562       & 14492$^{+225}_{-286}$ & 913$^{+189}_{-96}$ & 21 & r$<$10 arcmin \\
\noalign{\smallskip}
\hline
\end{tabular}
\end{flushleft}
\end{table}
%-------------------------------------------------------------------------------

%-------------------------------------------------------------------------------
\subsection{Luminosity functions}

In order to estimate the cluster luminosity function it is necessary to
correct the number of galaxies for the foreground and background contamination.
In general this is done assuming a universal number--counts relation, as done 
for example by Colless (1989) for a sample of 14 rich clusters and by Metcalfe 
et al. (1994) for A3558.
This approach is not correct for our clusters, because of the existence of the 
underlying supercluster, the presence of the Great Attractor overdensity at 
$\sim 4000$ km/s and of other structures behind the Shapley Concentration (see 
Bardelli et al. 1997) in approximately the same direction of the A3558 
complex: therefore we chose to estimate a local background, 
extracting counts from regions without obvious overdensities in the same
UKSTJ plate which contains the A3558 complex ($\# 444$),
obtaining a total control area of $\sim 10$ square degrees. 
Comparing these counts with those obtained from a reference region of $\sim 
140$ square degrees toward the South Galactic Pole (from the Edinburgh--Durham 
Southern Galaxy Catalogue, Heydon--Dumbleton et al. 1989) we found, as 
expected, that they present an excess at all magnitudes.
Such an excess corresponds to an average factor of $\sim 1.5$ at 
$17.5<b_J<19.5$.

Assuming that the luminosity function is well described by a Schechter (1976)
form 
\begin{equation}
\phi (L)dL= \phi^* \left( {{L}\over{L^*}}\right)^{\alpha} e^{-L/L^*} 
d \left({{L}\over{L^*}}\right)
\end{equation}
the probability of seeing a galaxy with $L_i>L_{min}$ is
\begin{equation}
p_i= \left({ {\phi(L_i)}\over { 
\int_{L_{min}}^{\infty} \phi(L_i) }}\right)^{w_i}
\end{equation}
where $L_{min}$ is the minimum absolute luminosity corresponding to the limiting
apparent magnitude of our survey ($b_J=19.5$) at the distance of each cluster.
The term $w_i$ is a weight which takes into account the background correction
and corresponds to the probability that the $i^{th}$ galaxy in the photometric 
sample belongs to the cluster.
In general, $w_i$ is a function of both magnitude $m$ and distance $r$ from
the cluster center and can be computed as
\begin{eqnarray}
w_i (m,r) = { N_{cl}(m,r) \over N_{tot}(m,r)  }= 
\nonumber \\
       =  { {\displaystyle{ [ N_{tot} (m) - N_{bck} (m) ] f(r) {\rm d}r } } 
         \over {\displaystyle{[ N_{tot} (m) - N_{bck} (m) ] f(r) {\rm d}r 
           + N_{bck} (m) { {2\pi r}\over{\pi r^2_{max}}} {\rm d}r } } }
\label{eq:pesomr}
\end{eqnarray}
where $N_{tot}$ is the number of galaxies in the projected cluster region and 
$N_{bck}$ is the number of ``background" galaxies expected in the same region
(from the estimate in the control area) and $N_{cl} = N_{tot} - N_{bck}$. 
The function $f(r)$ represents the ratio 
\begin{equation}
f(r) = { {\displaystyle { 2\pi r I(r) }} \over
         {\displaystyle {\int_0^{r_{max}} 2\pi r I(r) {\rm d} r }} }
\end{equation}
where $I(r)$ is the surface density of galaxies (see next Section) and $r_{max}$
is the maximum distance from the cluster center. 

If we neglect the radial dependence, as usually done in the literature,
eq.(\ref{eq:pesomr}) reduces to
\begin{equation}
w_i= 1- { N_{bck}(m) \over N_{tot}(m) }
\label{eq:pesom}
\end{equation}
The numbers $N_{bck}$ and $N_{tot}$ were computed in bins of 0.2 magnitudes,
for both eq.(\ref{eq:pesomr}) and eq.(\ref{eq:pesom}).
The parameters reported in the following were derived using 
eq.(\ref{eq:pesomr}): however the parameters obtained with 
eq.(\ref{eq:pesom}) are always consistent with them within $1 \sigma$ errors.

The parameters of the fitting function are derived through a maximum
likelihood technique, minimizing the product of the $p_i$ taking into account 
the weights. This method is the same as that applied by Zucca et al. (1994),
estimating the luminosity function when the sample is affected
by incompleteness. In the incompleteness case, 
the weights are greater than one in order to correct for the objects missed
or not observed in a given magnitude bin. 
On the contrary, in our case the weights are smaller than one 
because, due to the presence of the background contamination, the sample
used to compute the luminosity function (i.e. the photometric catalogue) 
contains also objects which do not belong to the clusters. 

Very recently, Lumsden et al. (1997) found that the COSMOS magnitude 
scale is not linear for bright objects ($b_J \la 17$), because of a lack 
of dynamic range of the measuring machine, and proposed a correction
for $15<b_J<21$ based on the comparison with CCD photometry. 
The corrected magnitudes are brighter than the original ones and 
this effect is at maximum $\sim 0.45$ mag at $b_J=15$. Therefore, 
when fitting the cluster luminosity functions, we adopted this correction
and we considered only galaxies with $b_J > 15$. 

Finally, the correction for galactic absorption has been applied using the 
extinction values reported by Burstein \& Heiles (1984): the values are 
$A_B=0.181$, 
$A_B=0.169$ and $A_B=0.189$ in the direction of the center of A3558, A3562 
and A3556, respectively.

The best fit parameters (with $1\sigma$ errors) of the Schechter form of the 
luminosity function of these three clusters are reported in Table 4 (cluster 
name in column 1, $\alpha$ and $M^*$ in column 2 and 3); the discussion of the
results is presented in Sect.4.

The parameters $M^*$ presented in the following have been derived including
the Lumsden et al. and the galactic absorption corrections, adopting a
distance modulus of 35.79 \hmagm. On the contrary,
when speaking about apparent magnitudes we always refer to the original
(i.e. without corrections) COSMOS magnitudes.

%-------------------------------------------------------------------------------
%%%%%%%%%%%%%%%%%%%%%%%%%%%%%%%%%%%%%%%
%TAB. 4 PARAMETRI DELLA FUNZIONE DI LUMINOSITA`
%%%%%%%%%%%%%%%%%%%%%%%%%%%%%%%%%%%%%%%
\setcounter{table}{3} 
\begin{table}
\caption[]{ Parameters of the cluster luminosity functions }
\begin{flushleft}
\begin{tabular}{llll}
\hline\noalign{\smallskip}
Name & $\alpha$ & $M^*$ & notes \\
\noalign{\smallskip}
\hline\noalign{\smallskip}
A3556 & -1.10$^{+0.32}_{-0.29}$ & -19.14$^{+0.50}_{-0.73}$ & $b_J>$15.5 \\
A3558 & -1.39$^{+0.12}_{-0.12}$ & -20.26$^{+0.39}_{-0.56}$ & $b_J>$15.0 \\
A3562 & -1.42$^{+0.19}_{-0.15}$ & -19.84$^{+0.46}_{-0.61}$ & $b_J>$15.0 \\
\noalign{\smallskip}
\hline
\end{tabular}
\end{flushleft}
\end{table}
%-------------------------------------------------------------------------------

%-------------------------------------------------------------------------------
\subsection{Density profiles}

For the fit of the radial density galaxy profiles, we assume a King law
of the form
\begin{equation}
I(r)= I_o \left( 1+ \left( {r \over {r_c}} \right)^2 \right)^{-\alpha}
\end{equation}
where $I(r)$ is the surface density of galaxies (in \hmendue or arcmin$^{-2}$) 
at distance $r$ from the cluster center, $r_c$ is the core radius (in \hmpc or
arcmin), $\alpha$ is the profile steepness (generally fixed to one) 
and $I_o$ is the normalization (in \hmendue or arcmin$^{-2}$). 
We minimize the likelihood function
\begin{equation}
{\cal L}=\prod_1^{N_{gal}}     {{ 2 \pi r_i \left[ I(r_i)+bck \right]} 
\over{\int_0^{r_{max}}
2 \pi r \left[ I(r)+bck \right] d r }}
\end{equation}
where $bck$ is the background (\hmendue or arcmin$^{-2}$), $r_i$ (\hmpc or
arcmin) is the radial distance of the i$^{th}$ galaxy from the cluster center
and $r_{max}$ is the maximum radial extension of the sample.

The results for A3556, A3558 and A3562 are reported in Table 5: column 1
lists the cluster name, column 2 gives $\alpha$, in column 3 there is $r_c$
(in arcmin), and columns 4 and 5 report $I_o$ and $bck$ (in arcmin$^{-2}$),
respectively.
The discussion of the various cases is presented in Sect.4.

%-------------------------------------------------------------------------------
%%%%%%%%%%%%%%%%%%%%%%%%%%%%%%%%%%%%%%%
%TAB. 5 PARAMETRI distribuzione radiale
%%%%%%%%%%%%%%%%%%%%%%%%%%%%%%%%%%%%%%%
\setcounter{table}{4} 
\begin{table}
\caption[]{ Parameters of the radial density profiles }
\begin{flushleft}
\begin{tabular}{llllll}
\hline\noalign{\smallskip}
Name & $\alpha$ & $r_c$    &       $I_0$     & bck             & notes \\
     &          & (arcmin) & (arcmin$^{-2}$) & (arcmin$^{-2}$) &       \\
\noalign{\smallskip}
\hline\noalign{\smallskip}
A3556 & 0.55 & 0.90 & 0.45 & 0.01 & $b_J \le 18$ \\
A3558 & 0.85 & 3.60 & 0.90 & 0.15 &          \\
A3562 & 0.70 & 3.10 & 0.70 & 0.10 &          \\
\noalign{\smallskip}
\hline
\end{tabular}
\end{flushleft}
\end{table}
%-------------------------------------------------------------------------------

Assuming spherical symmetry and that the orbits are isotropic and the velocity 
dispersion is not a function of the distance from the cluster center 
(i.e. the cluster is isothermal), it is possible to have an 
estimate of the total mass inside the radius $r$ using the formula 
(Binney \& Tremaine 1987)
\begin{equation}
M(<r)= - { {r \sigma^2} \over G} { {{\rm d}\ln \rho}\over {{\rm d}\ln r}}
\label{eq:massa}
\end{equation}
where $\rho$ is the spatial (i.e. tri--dimensional) distribution of galaxies,
$\sigma$ the velocity dispersion of the cluster and $M(<r)$ is the
total mass inside the radius $r$. 
Note that the hypothesis that the cluster is isothermal is supported 
(at least for A3558) by
the fact that the distributions of both hot gas temperature (see Paper II) 
and observed velocity dispersion (see Paper I) do not strongly vary with
the distance from the cluster center. 

The spatial distribution $\rho$ is derived from the galaxy surface density 
distribution, through the standard Abel inversion, and it results 
\begin{equation}
\rho= \rho_o \left( 1+ \left( {r \over {r_c}} \right)^2 
\right)^{-\alpha-0.5}
\end{equation}
where $\rho_o$ is the central density (see The \& White 1986).

%-------------------------------------------------------------------------------
\section{Results}

%-------------------------------------------------------------------------------
\subsection{A3556}

A3556 is a poor cluster (richness class 0), classified in the 
Bautz--Morgan class I, although it is dominated by two giant ellipticals
with $b_J=14.32$ and $b_J=14.49$ at $v=14074$ km/s and $v=14459$ km/s,
respectively. 

Previous determinations of the dynamical
parameters are from Paper I (\vmed $=14407\pm 89$ km/s, $\sigma=554 \pm 47$
km/s, using 48 galaxies), from Quintana et al. (1995; \vmed $=14167\pm 116$
km/s, $\sigma=538^{+98}_{-64}$ km/s, based on 25 galaxies) and Stein
(1997; \vmed $=14574\pm 86$ km/s, $\sigma=459 \pm 44$ km/s, based on 30 
galaxies). 
From these values it is clear that, although the velocity dispersion 
determinations are in agreement, the mean velocity estimates are not consistent 
within the quoted errors: the reason of this inconsistency is clarified
using our new sample.

The angular separation between the centers of A3556 and A3558 is only 
$\sim 50$ arcmin,
while the Abell radius at the distance of the clusters is $\sim 36$ arcmin.
Therefore, if the angular extension of the velocity sample around the cluster
center is comparable with these sizes, the contamination by objects from
the nearby cluster could be important. 
For this reason, we decided to consider only objects within $22$ arcmin from 
the cluster center in the analysis of the dynamical parameters: this value 
was chosen on the basis of a visual inspection of both the isodensity 
contours and the wedge diagram. 

%-------------------------------------------------------------------------------
% FIGURE 4. 
\begin{figure}
\epsfysize=8.5cm
\epsfbox{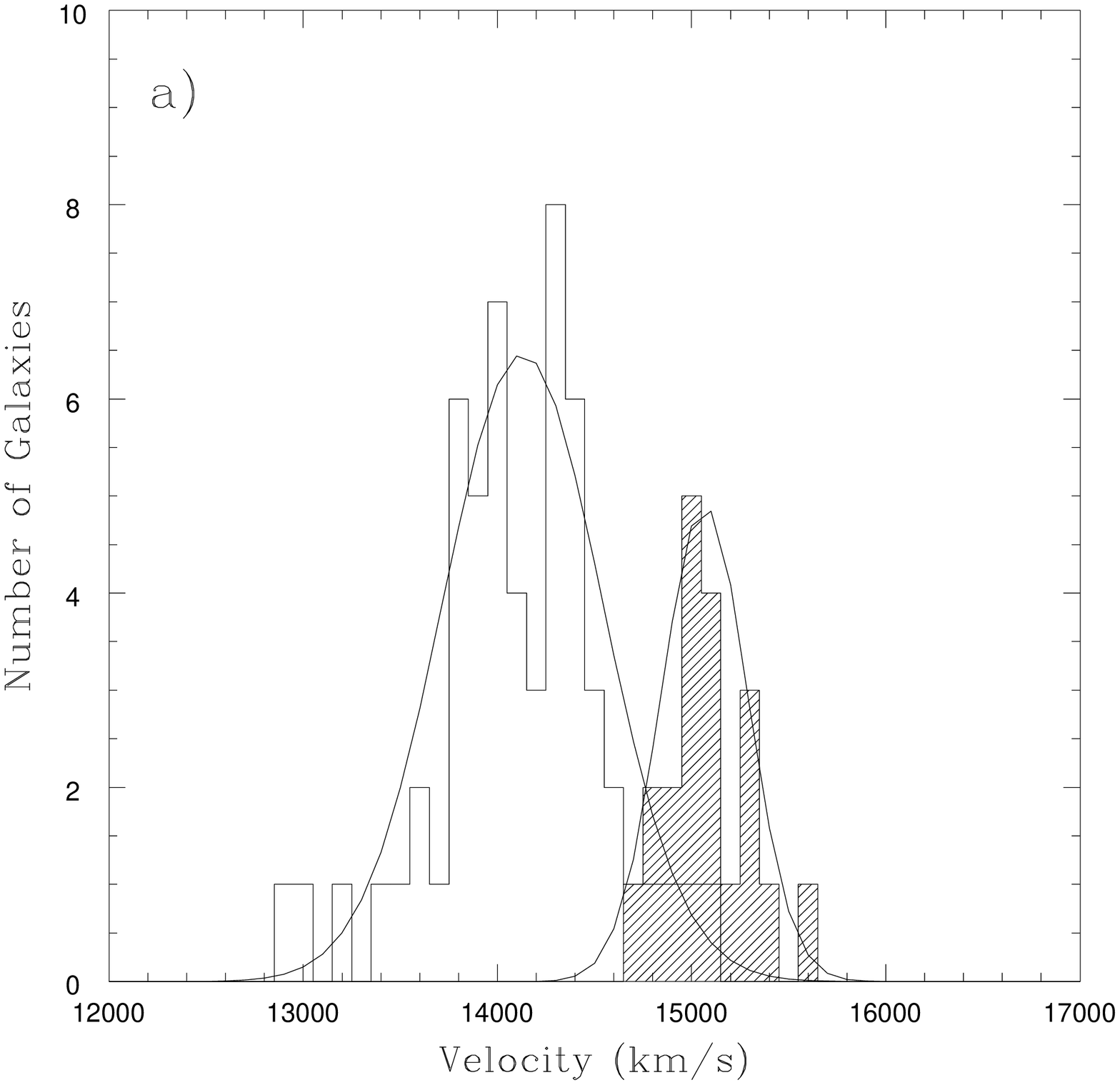}
\epsfysize=8.5cm
\epsfbox{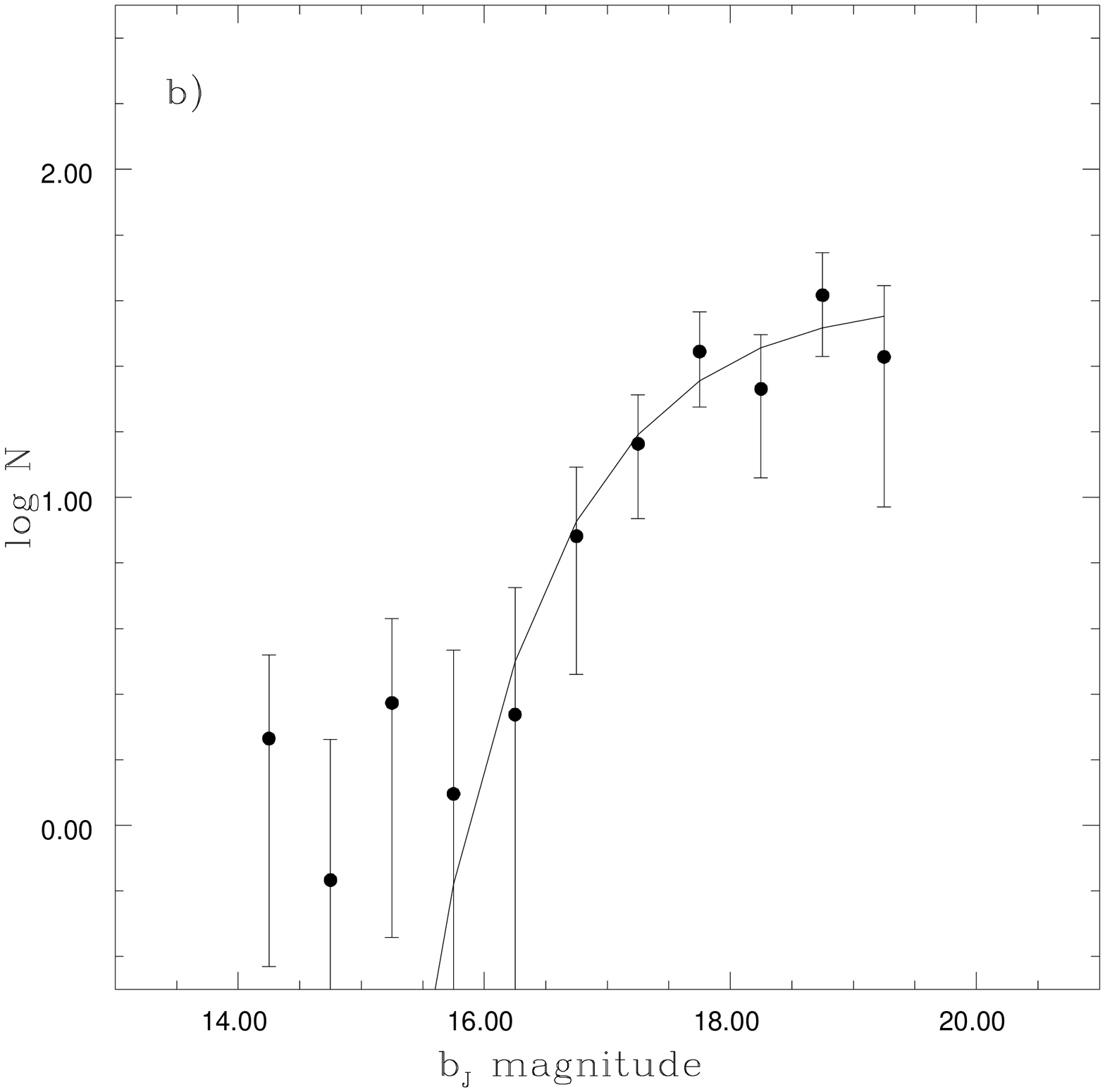}
\caption[]{
a) Redshift distribution for the galaxies within $22$ arcmin from
the center of A3556, divided in the two components.
Open histogram corresponds to the members of a 
group with \vmed $=14130^{+42}_{-74}$ km/s and $\sigma=411_{-29}^{+76}$ km/s,
while shadowed histogram corresponds to members of a group 
with \vmed $=15066^{+58}_{-43}$ km/s and $\sigma=222_{-59}^{+35}$ km/s.
\parn
b) Luminosity function of A3556. The fitted parameters are
$\alpha=-1.10^{+0.32}_{-0.29}$ and $M^*=-19.14_{-0.73}^{+0.50}$, after
having excluded the objects brighter than $b_J=15.5$. }
\label{fig:a3556}
\end{figure}
%-------------------------------------------------------------------------------
\setcounter{figure}{3} 
\begin{figure}
\epsfysize=8.5cm
\epsfbox{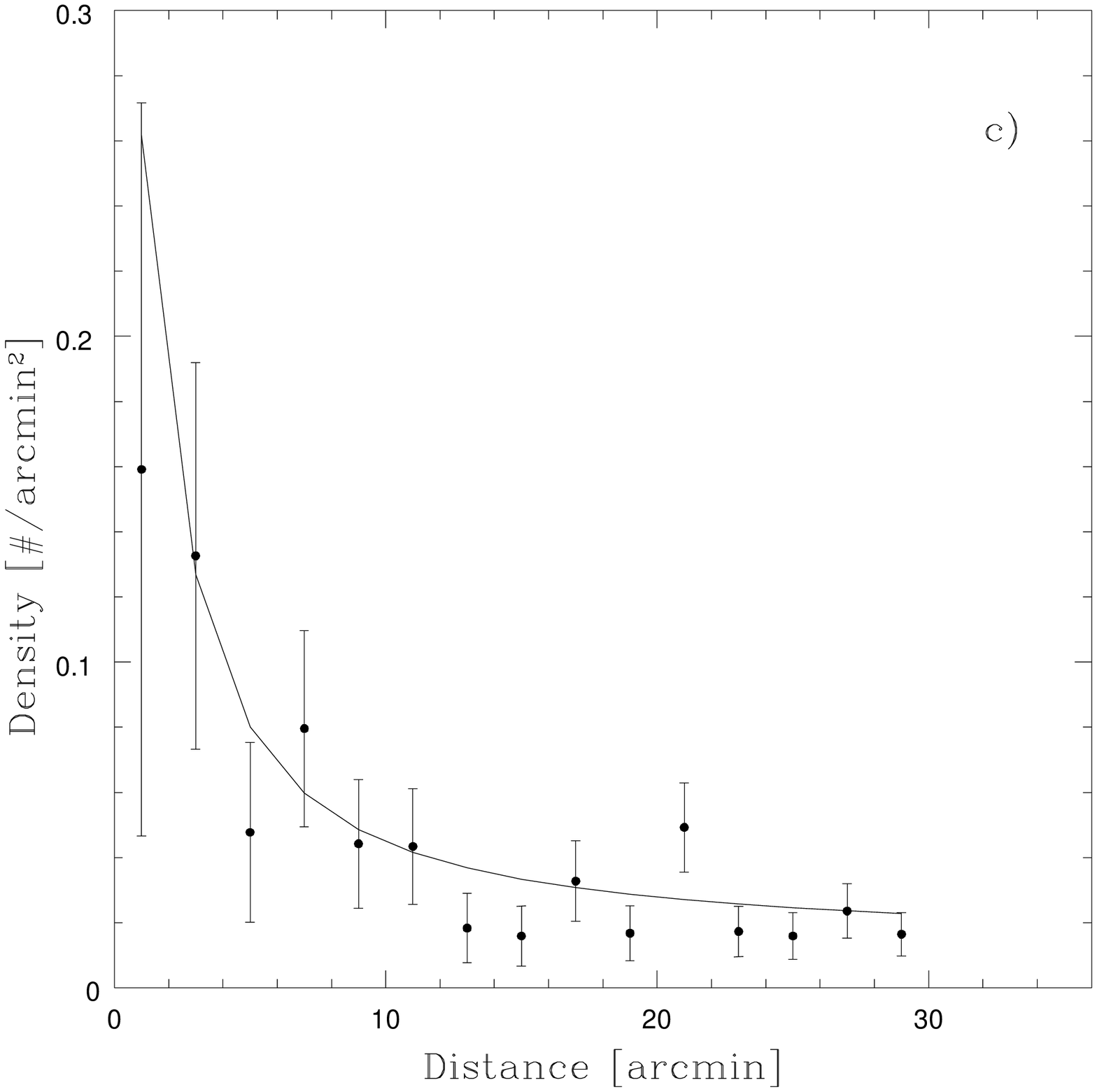}
\epsfysize=8.5cm
\epsfbox{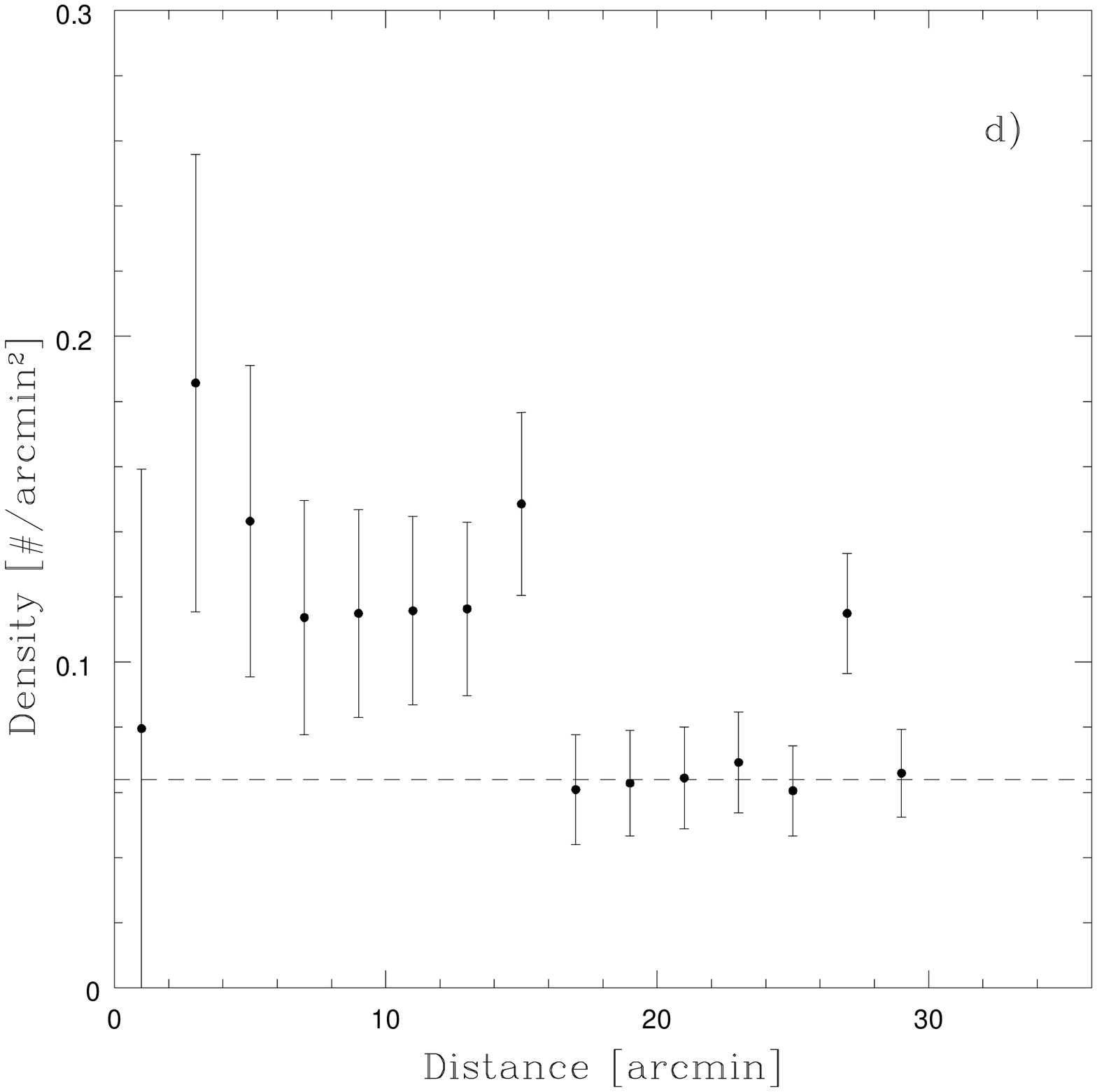}
\caption[]{
c) Galaxy density distribution in A3556, as a function of the distance
from the cluster center; only galaxies with $b_J \le 18$ are considered. 
No background subtraction has been applied in this figure.
\parn
d) Same as panel c), but considering objects with $18<b_J \le 19.5$. 
The dashed line represents the background as estimated in the control area
in the same magnitude range. 
} 
\label{fig:a3556prof}
\end{figure}
%-------------------------------------------------------------------------------

Using the new sample, the estimated parameters on the basis of $79$ 
velocities are 
\vmed $=14357_{-76}^{+76}$ km/s and $\sigma=643_{-43}^{+53}$ km/s. However, 
from the redshift histogram shown in Figure \ref{fig:a3556}a, it is clear 
that the distribution presents two distinct peaks:
applying the KMM test we found that the heteroscedastic bimodality  is
significant and the single Gaussian distribution is rejected at a 
confidence level of $97.5\%$. The biweight estimators of the position and
the scale for the two distributions are \vmed$_1 =14130^{+42}_{-74}$ km/s,
$\sigma_1 =411^{+76}_{-29}$ km/s and \vmed$_2 =15066^{+58}_{-43}$ km/s,
$\sigma_2 =222^{+59}_{-35}$ respectively. The number of objects assigned to
the two peaks is $56$ to the lower velocity clump and $23$ to the other one.
Note that the mean velocity determination of Quintana et al. (1995)
is well consistent with the first substructure: given the limited number
of redshifts used by these authors ($25$), their sample could be dominated
by this clump. A KS analysis of the velocity distribution between bright
($b_J\le 18$) and faint ($b_J>18$) galaxies excludes a luminosity segregation
between the two groups at a confidence level of $30\%$. Also the spatial
distribution on the plane of the sky does not present significant differences. 

The luminosity function of A3556, which is reported in Figure \ref{fig:a3556}b, 
is different from those of the other two clusters 
(see below), because it is rather flat for magnitudes brighter than $\sim 16$,
forming a ``plateau".
This excess of bright galaxies is confirmed by the fact that the number of 
galaxies in A3556 with $b_J< 15.5$ is the same as in A3558 (6 objects within 
one Abell radius), but its total number is only $60\%$ of the latter 
cluster. 
Considering only galaxies in the magnitude range $[15.5-19.5]$, we found 
$\alpha=-1.10^{+0.32}_{-0.29}$ and $M^*=-19.14_{-0.73}^{+0.50}$, with an 
effective (i.e. background subtracted) number of objects of $151$. 
These values are significantly different from the ``typical" parameters found 
for clusters: however, the observed luminosity function of this cluster 
could be a complex combination of two components (as suggested by the 
bimodality in the velocity distribution), whose single contributions are not 
easily separable.

Venturi et al. (1997) found that all the galaxies belonging to the
plateau of the luminosity function are radio loud, suggesting a possible
connection between the shape of the bright part of the luminosity function and
the enhanced probability for a galaxy to have radio emission.
Moreover, the presence of a remarkable low surface brightness extended radio 
source (dubbed J1324-3138 by Venturi et al. 1997) is revealed. It seems 
to be associated
with a $b_J=15.6$, $v=15142$ km/s optical galaxy. Venturi et al. (1997)
concluded that, given its steep spectrum, this source could be a narrow
angle tail radio galaxy at the latest stage of its life. 
The fact that J1324-3138 belongs to the higher velocity group, which on the 
basis of the velocity dispersion could be considered the least massive 
component and thus falling toward the main component of A3556, suggests the 
presence of significant ram pressure acting on the radio source due to the 
interactions between intracluster media.
This effect probably accounts for the evolution of radio sources in the diffuse 
emissions typically found in merging clusters (Feretti \& Giovannini 1996)
and could be responsible for the properties of  J1324-3138.

The galaxy density profile for this cluster is not well fitted by a King law.
This fact is probably due to the different distribution of bright and faint
galaxies, as shown in Figures \ref{fig:a3556prof}c and \ref{fig:a3556prof}d,
where the profiles for galaxies with $b_J \le 18$ and $b_J > 18$ are reported.
The bright galaxies distribution appears quite centrally peaked, as confirmed
by the parameters of the fit: we find $I_o=0.45$ arcmin$^{-2}$ (corresponding 
to $258$ \hmendue), $r_c=0.90$ arcmin ($0.04$ \hmpc), $\alpha=0.55$ and 
$bck=0.01$ arcmin$^{-2}$ ($7.5$ \hmendue). The fitted background is in good 
agreement with the value derived from the control area, in the same magnitude
range ($bck=0.011$ arcmin$^{-2}$). On the contrary, the distribution of
faint galaxies appears quite flat up to a radius of 15 arcmin, while it is
consistent with the background density (dashed line in Figure 4d) for larger
radii: this distribution is not well fitted with a simple King profile.
For this reason, the parameters of the luminosity function given above were 
derived computing $w_i$ following eq.(\ref{eq:pesomr}) for $b_J \le 18$ and 
eq.(\ref{eq:pesom}) for $b_J > 18$.

The complex behaviour of the density profile could be related to the presence
of the two substructures found in the dynamical analysis: however, it is 
not possible to separate the contribution of these two components and 
therefore to properly estimate the cluster mass.

%-------------------------------------------------------------------------------
\subsection{A3558}

A3558, alias Shapley 8 or SC 1325--11, is a Bautz--Morgan type I cluster,
classified by ACO in the richness class 4, and could be regarded as the
dominant component of the cluster complex. In literature there is a large
number of galaxy redshifts taken by Metcalfe et al. (1987), Teague et al. 
(1990), Stein (1996) and by us (Paper I). 
We estimated a mean velocity and a velocity dispersion of \vmed $=14242\pm 80$ 
km/s and $\sigma=986\pm 60$ km/s, within $1077$ arcsec (corresponding to
$\sim 0.5$ Abell radii) from the cluster center. In Paper I we showed that for 
larger distances the contamination by objects belonging to the nearby groups 
(SC 1327--312 and SC 1329--314) becomes not negligible. 

In Paper II we studied a deep ($\sim 30$ ksec) ROSAT PSPC X--ray observation
of this cluster in the $[0.1-2.4]$ keV band. This cluster results to be rather 
cold ($KT=3.25$ keV), with an X--ray total luminosity of $1.1\times 10^{44}$
h$^{-2}$ erg s$^{-1}$ in the $[0.5-2.0]$ keV band. The dynamical mass,
estimated on the basis of the hydrostatic equilibrium, is $3.1\times 10^{14}$
h$^{-1}$ M$_{\odot}$ inside one Abell radius and results to be a factor 4 
lower than the virial mass estimate of Metcalfe et al. (1994). 
Given the X--ray luminosity and the dynamical mass, it is clear that the
optical richness of A3558 is overestimated (Paper II, Breen et al. 1994):
Metcalfe et al. (1994) estimated a corrected richness class of 2.
The explanation of this overestimate is probably the contamination
due to nearby groups in this region.

%-------------------------------------------------------------------------------
% FIGURE 5. 
\begin{figure}
\epsfysize=10cm
\epsfbox{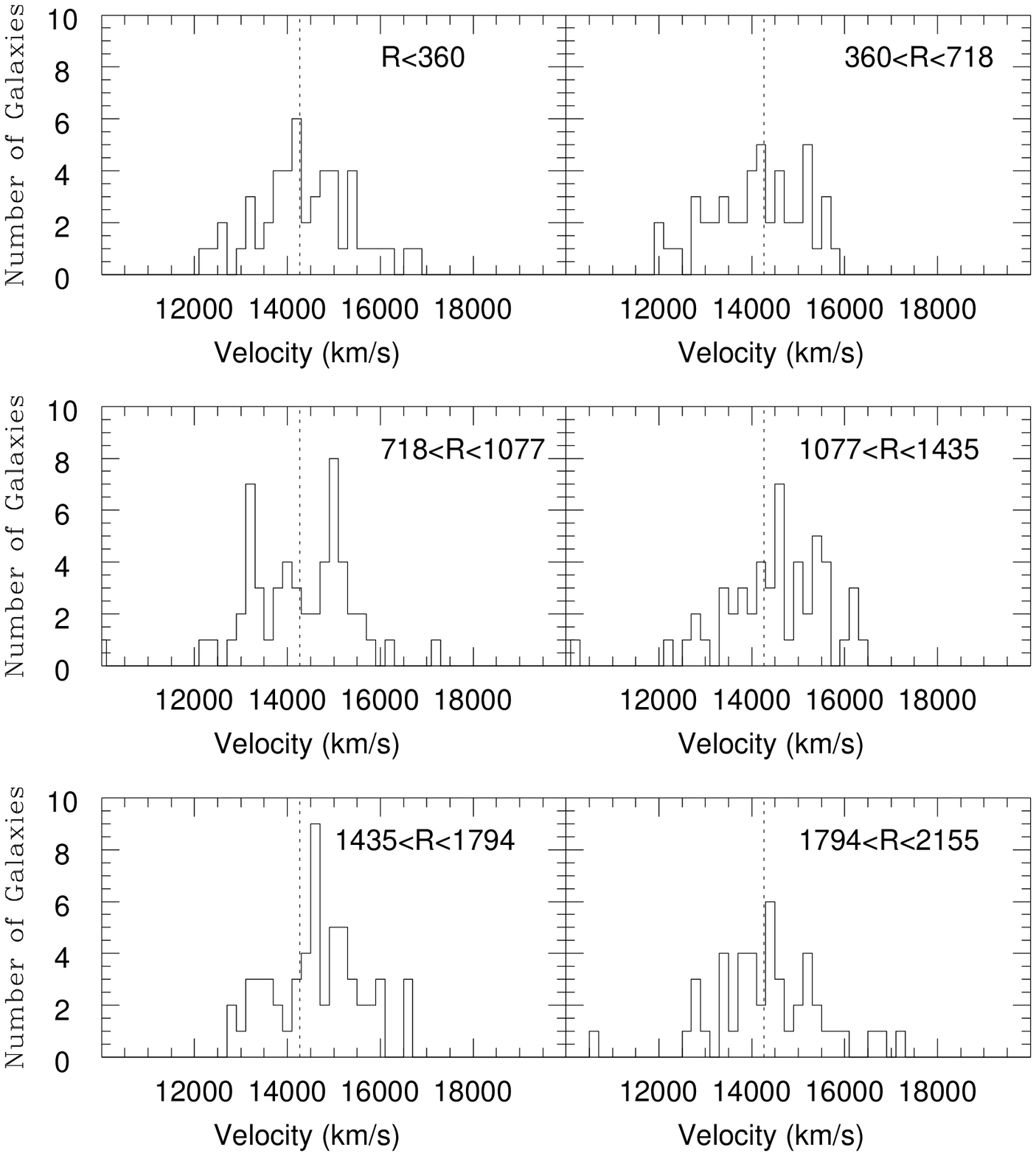}
\caption[]{Redshift distribution of A3558 in various shells centered on the 
cluster center. In particular note, in the $[718 - 1077] $ arcsec shell,
the presence of two peaks at $v\sim 13000$ km/s and $v\sim 15000$ km/s. The
dashed line represents the mean velocity of $14262$ km/s, determined 
within a region of $\sim 0.5$ Abell radii.} 
\label{fig:a3558his}
\end{figure}
%-------------------------------------------------------------------------------

%-------------------------------------------------------------------------------
% FIGURE 6. 
\begin{figure}
\epsfysize=8.5cm
\epsfbox{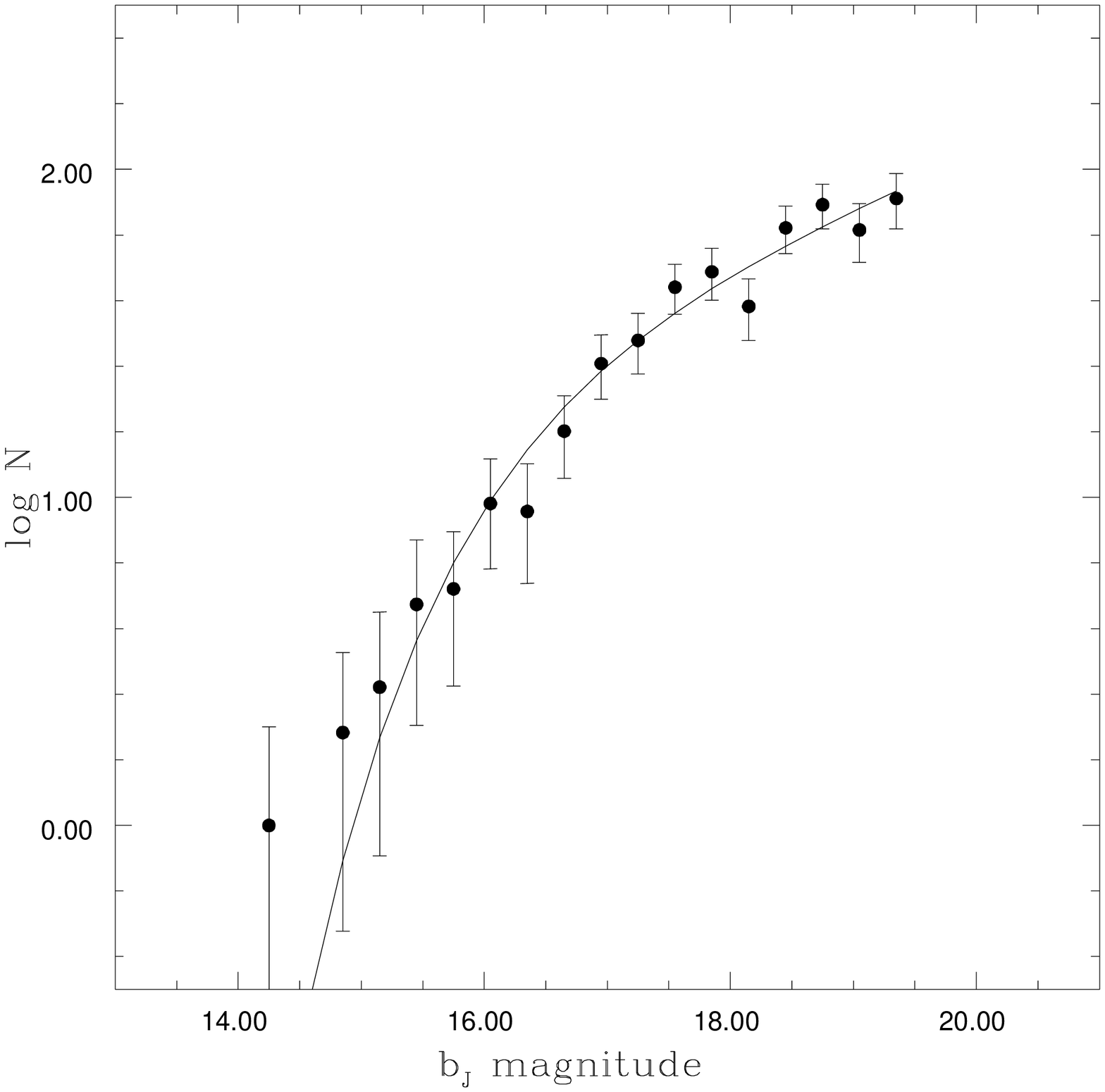}
\caption[]{Luminosity function of A3558. The fitted parameters are
$\alpha = -1.39^{+0.12}_{-0.12}$ and $M^*=-20.26_{-0.56}^{+0.39}$. The effective
number of galaxies is 525.}
\label{fig:a3558lf}
\end{figure}
%-------------------------------------------------------------------------------

Very recently, Markevitch \& Vikhlinin (1997), studying a 17 ksec
image of A3558 taken with the GIS detector of ASCA, estimated a temperature
of $5.5$ keV. On the basis of the gas temperature -- velocity dispersion 
relation of Lubin \& Bahcall (1993)
\begin{equation}
 \sigma=332 \left( KT \right)^{0.6}   
\label{eq:lubin}
\end{equation}
the higher ASCA temperature appears to be more consistent with the galaxy 
velocity dispersion than the ROSAT one (which corresponds to $\sigma \sim
700$ km/s).
On the other hand, Bird (1994) and Girardi et al. (1997), applying various tests
for substructure detection, found that the velocity dispersion
of A3558 is significantly affected by subclumps. After having 
corrected the sample for this effect, they determine a velocity dispersion
of $781^{+178}_{-187}$ km/s and $735^{+49}_{-41}$ km/s, respectively.

The discrepancy between the two temperature values could indicate
the existence of a calibration problem in one of the two satellites.
However, this situation is very similar to the case of A2255 (Feretti et al. 
1997), for which the ROSAT temperature ($3.5$ keV) is significantly lower than 
that obtained by Einstein ($7.3$ keV) in a slightly higher energy window, 
and of Hydra A (Ikebe et al. 1997), for which $T_{ROSAT} < T_{ASCA}$. 
The authors claim that this discrepancy could be the result of several
co--existing components at different temperatures.

%-------------------------------------------------------------------------------
% FIGURE 7. 
\begin{figure}
\epsfysize=8.5cm
\epsfbox{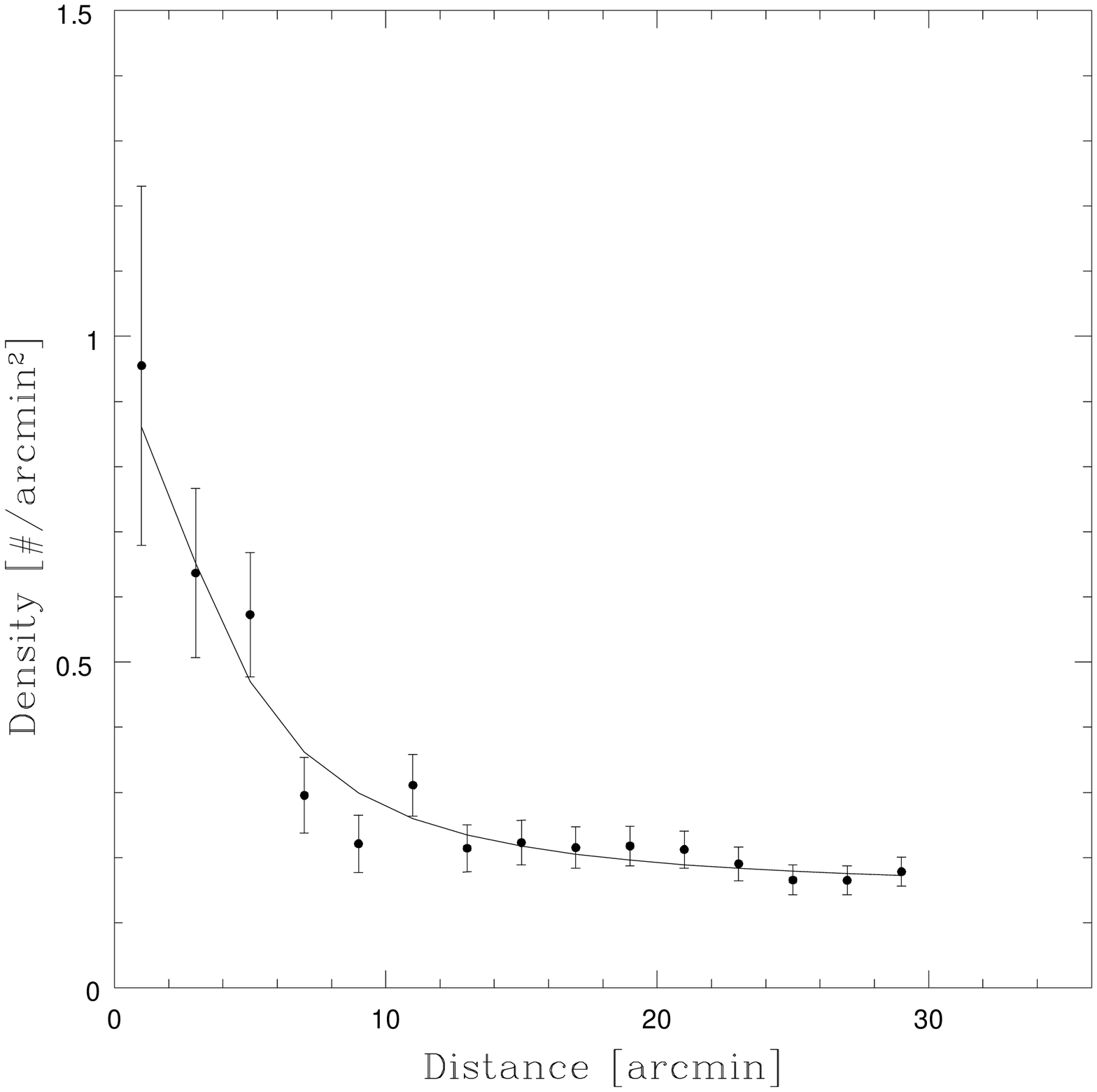}
\caption[]{
Galaxy density distribution in A3558, as a function of the distance
from the cluster center. No background subtraction has been applied 
in this figure. } 
\label{fig:a3558prof}
\end{figure}
%-------------------------------------------------------------------------------

Adding the new redshift data to the catalogue, we estimate a global mean 
velocity of \vmed $=14403^{+60}_{-55}$ km/s
and a velocity dispersion of $\sigma=996^{+40}_{-36}$ km/s within one Abell 
radius and using 307 galaxies. However, the considered region
contains SC 1327--312 and its boundaries are very nearby to SC 1329--313
and A3556 and could be contaminated by galaxies of these clusters.
For this reason, we restricted the analysis to $0.5$ Abell radii finding 
\vmed $=14262^{+75}_{-82}$ km/s and $\sigma=992^{+85}_{-60}$,determined with 
155 objects. Both these values are in agreement with those found in Paper I.

Metcalfe et al. (1994), on the basis of 88 redshifts, found a difference
in velocity dispersion between galaxies fainter and brighter than $m_{B}=17$ 
and belonging to the ``red sequence" characterizing the cluster. 
Using all galaxies within $0.5$ Abell radii, without any selection in  
color,
we find the same effect, but at a marginal (1.7 sigma) significance level.
In fact, the 31 galaxies brighter than $b_J=17$ have 
\vmed $=14309^{+285}_{-219}$ km/s and $\sigma=1125^{+139}_{-79}$ km/s, while
the fainter ones (117 objects) have \vmed $=14260^{+100}_{-67}$ km/s and 
$\sigma=946^{+70}_{-80}$ km/s.
At face value, this result is opposite to that of Biviano et al. (1992) who
found, using a sample of 68 clusters, that the brightest members have a 
lower velocity dispersion with respect to the faintest ones.
However, from the KMM analysis we find that the higher velocity dispersion
of bright galaxies 
may be due to the superposition of two separate groups (homoscedastic 
bimodality significant at $96 \%$).
If we assume that the system is described by two Gaussians, the biweight means 
and velocity dispersions are \vmed $=12993^{+129}_{-119}$ km/s and 
$\sigma=310^{+119}_{-92}$ km/s ($N_{gal}=11$) and \vmed $=14871^{+144}_{-145}$
km/s and $\sigma=455^{+51}_{-55}$ km/s ($N_{gal}=20$).
These two groups do not appear to be separated on the plane of the sky.
We consider this bimodality, which can not be discussed further due to the 
small number of galaxies, only as an indication of the complex velocity 
distribution of bright galaxies in A3558.

The complicated dynamical state of the whole cluster is well illustrated
by Figure \ref{fig:a3558his}, where we show the velocity histograms in the six 
radial distance bins defined in Paper I. 
The dashed line is the mean velocity as estimated in the inner 1077 arcmin 
($14262$ km/s). The dynamical parameters 
calculated in these velocity intervals are essentially the same as those
of Table 5 in Paper I: the mean velocity is constant inside $0.5$ Abell radii 
and increases in the three most external bins. 
In five out of six bins the velocity histograms are reasonably symmetric and 
featureless. The only exception is the third bin ($718<R<1077$ arcsec), 
where two peaks at $13200$ and $15000$ km/s are clearly visible.
The bimodality is significant at the $98\%$ level, giving 
\vmed$_1 =13457_{-114}^{+115}$ km/s, $\sigma_1=470_{-68}^{+88}$ km/s and 
\vmed$_2 =15011_{-91}^{+77}$ km/s, $\sigma_2=349_{-59}^{+71}$ km/s, 
with $N_1=25$ and $N_2=27$ galaxies, respectively. 
Note that the structure at higher velocity has parameters which
are consistent within 1 sigma with the farther component of the distribution
of the bright galaxies discussed above.

An analysis of the isodensity contours (see for example Figure 
14b of Paper I), shows the presence of an excess of galaxies
in this third shell, southern of the center of A3558. 
This substructure is revealed by the West et al. (1988) test, which
measures the degree of symmetry of the bi--dimensional distribution with 
respect to the cluster center. Most of the galaxies in the lower velocity peak
reside in this substructure: moreover, in this density excess there
are two galaxies detected in the ROSAT X--ray band (\#6 and \#9 in Paper II)
with $v=13186$ km/s, $b_J=15.1$, and $v=12801$ km/s, $b_J=15.8$,
respectively. 

In order to detect other substructures, we applied the Dressler \& Shectman
(1988) $\Delta$ test, which uses both positional and velocity information.
These authors suggest a bootstrap procedure in order to estimate the presence
of significant substructures in a cluster: following this procedure we found 
that A3558 shows evidence of substructures at more than $3 \sigma$ level,
and the most significant signal is due to a condensation centered at 
$\alpha(2000) \sim 13^h 29^m 16^s$ and $\delta(2000) \sim -31^o 18' 45"$,
well in agreement with the position of the structure found by Girardi et al. 
(1997) and $\sim 10$ arcmin away from that found by Bird (1994). This 
substructure is in the quadrant of the cluster where Markevitch \& 
Vikhlinin (1997), in their analysis of an ASCA X--ray map of A3558, found 
a significant increase of the hot gas temperature.
The detailed analysis of the substructures in A3558 will be presented 
elsewhere (Bardelli et al., in preparation).

The parameters of the Schechter fit to the luminosity function are 
$\alpha = -1.39^{+0.12}_{-0.12}$ and $M^*=-20.26_{-0.56}^{+0.39}$, when galaxies
within one Abell radius are considered (Figure \ref{fig:a3558lf}). 
The effective number of objects belonging to the cluster is $525$, i.e. 
$68\%$ of the total. 
This result does not significantly change when considering a smaller radius. 
These values are in agreement with those found by Lumsden et al. (1997),
who estimated an average luminosity function with $\alpha=-1.22\pm 0.04$ and 
$M^*=-20.16\pm 0.02$, using a sample of 22 clusters, obtained from the 
Edinburgh--Durham Southern Galaxy Catalogue (derived from COSMOS scans)
and therefore directly comparable with our data.
These values are also consistent with those found by Colless (1989) for
14 APM clusters ($\alpha=-1.21$ and $M^*=-20.04$).
This agreement indicates that the magnitude distribution of galaxies in A3558
is not strongly modified by the interactions between this cluster and the 
nearby ones.

On the contrary, Metcalfe et al. (1994), on 
the basis of scans of plates in the $B$ band, found a flatter 
luminosity function ($\alpha=-1.0$, $M^*_B =-19.50$) for A3558. 
This discrepancy arises from the different galaxy counts in the two photometric 
samples, but its origin is unclear: it could be due to incompleteness 
of one of the two catalogues at faint magnitudes, or to systematic photometric 
errors, or to problems in the plate quality. 

For what concerns the radial density profile, as already noted by Metcalfe et 
al. (1994), the parameters are poorly constrained. The minimum of the
likelihood function is not well defined and this leads to very large
uncertainties on the parameters. Our best estimates are $I_o=0.90$ 
arcmin$^{-2}$ (corresponding to $459$ \hmendue), $r_c=3.6$ arcmin 
($0.15$ \hmpc), $\alpha=0.85$ and $bck=0.15$ arcmin$^{-2}$ ($86$ \hmendue)
(Figure \ref{fig:a3558prof}). Note that the fitted background is a factor 
$\sim 2$ higher than the value derived from the control area ($bck=0.076$ 
arcmin$^{-2}$): this is probably due to the fact that A3558 is embedded
in the center of a high density structure. 

The comparison with the results of
Metcalfe et al. (1994) is not straightforward, because $I_o$ and $bck$
depend on the different photometric band: moreover, their parameters
were obtained by fixing the slope or the background. However, our slope and
core radius are consistent at $\sim 10\%$ with those of case (1) for the red
sequence galaxies of Metcalfe et al. (see their Table 4).

From the radial distribution of galaxies we estimated the mass of the cluster 
(eq.\ref{eq:massa}): using a velocity dispersion of $\sigma=992$ km/s we found 
$M(< 0.75$\hmpc$)= 4.5 \times 10^{14}$ h$^{-1}$ M$_{\odot}$ 
and $M(< 1.5$\hmpc$)= 9.2 \times 10^{14}$ h$^{-1}$ M$_{\odot}$, 
$\sim 1.3$ times lower than the estimates of Biviano et al. (1993) 
[$M(< 0.75$\hmpc$)= 6 \times 10^{14}$ h$^{-1}$ M$_{\odot}$] and
Metcalfe et al. (1994) [$M(< 1.5$\hmpc$)= 1.2 \times 10^{15}$ h$^{-1}$ 
M$_{\odot}$], obtained applying the virial theorem.
On the other hand Dantas et al. (1997), after having removed a
substructure, estimated $M(<0.75$\hmpc$)= 3.40 
\times 10^{14}$ h$^{-1}$ M$_{\odot}$, under the hypothesis that galaxies
orbit around a common central potential.

These differences can be a consequence of the different hypotheses on which 
the mass estimate methods are based (see e.g. the discussion in
Bahcall \& Tremaine 1981) and/or of the different choices for the 
cluster parameters, as f.i. the velocity dispersion. Our derived mass is 
factor $\sim 3$ or $\sim 2$ greater than the X--ray determination, using a 
gas temperature of $3.25$ keV or $5.5$ keV, respectively.

%-------------------------------------------------------------------------------
% FIGURE 8. 
\begin{figure}
\epsfysize=8.5cm
\epsfbox{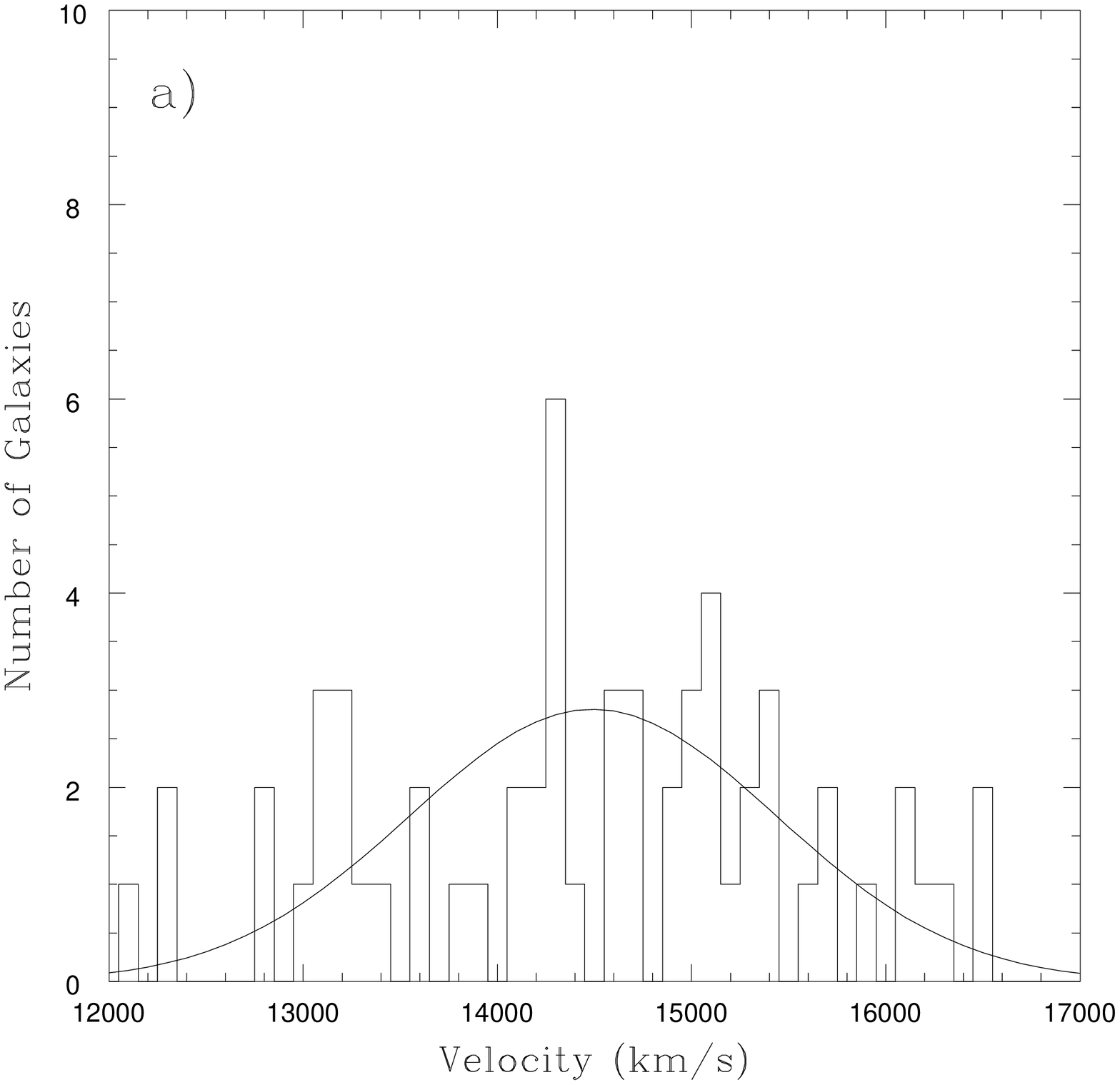}
\epsfysize=8.5cm
\epsfbox{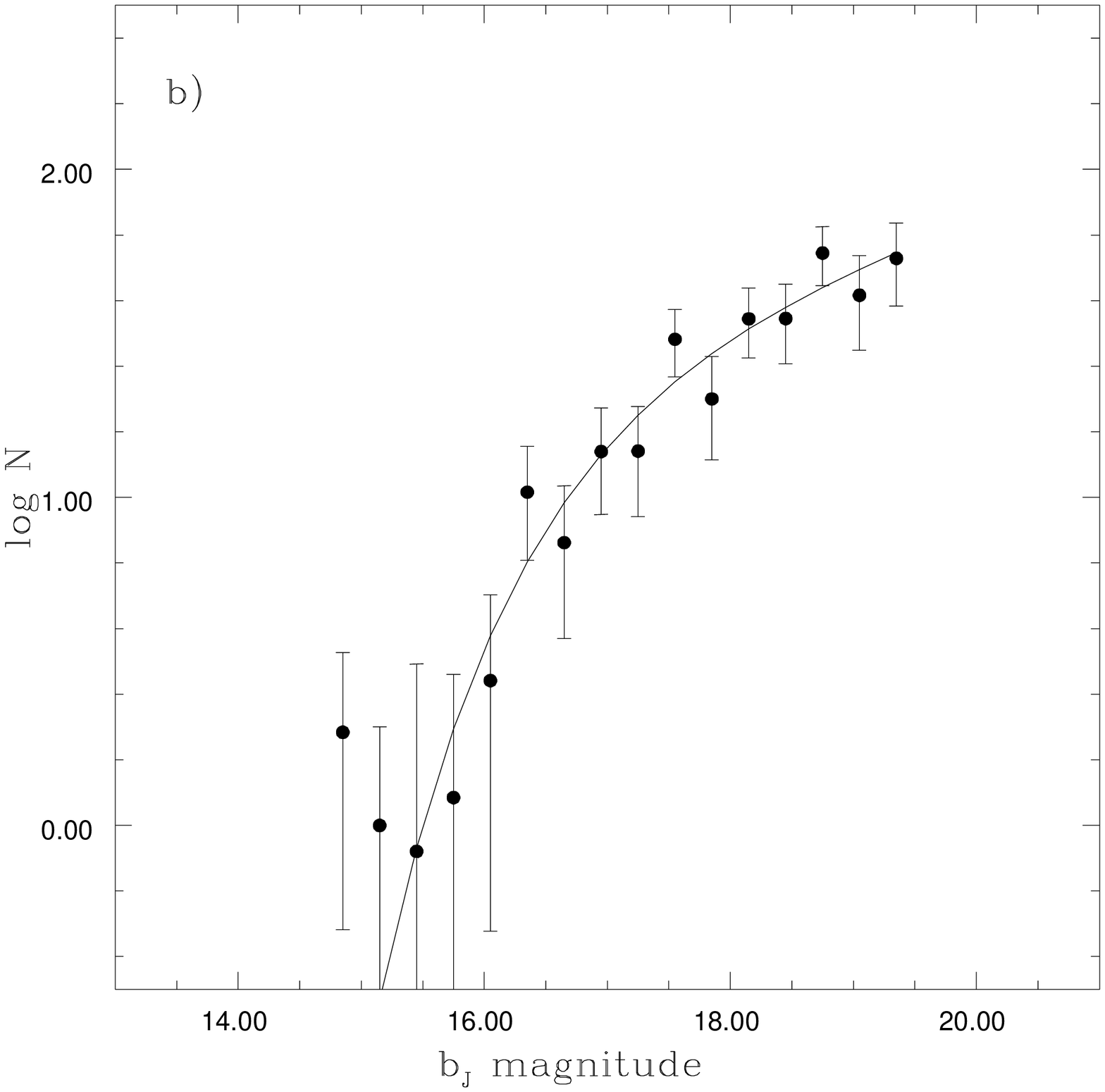}
\caption[]{ a) Redshift distribution of A3562. The Gaussian has parameters
\vmed $=14492^{+225}_{-286}$ km/s and $\sigma=913_{-96}^{+189}$ km/s,
obtained from a sample of 21 galaxies within 10 arcmin from the cluster
center. 
\parn
b) Luminosity function of A3562. The fitted parameters are 
$\alpha=-1.42_{-0.15}^{+0.19}$ and $M^*=-19.84_{-0.61}^{+0.46}$.
}
\label{fig:a3562}
\end{figure}
%-------------------------------------------------------------------------------
\setcounter{figure}{7} 
\begin{figure}
\epsfysize=8.5cm
\epsfbox{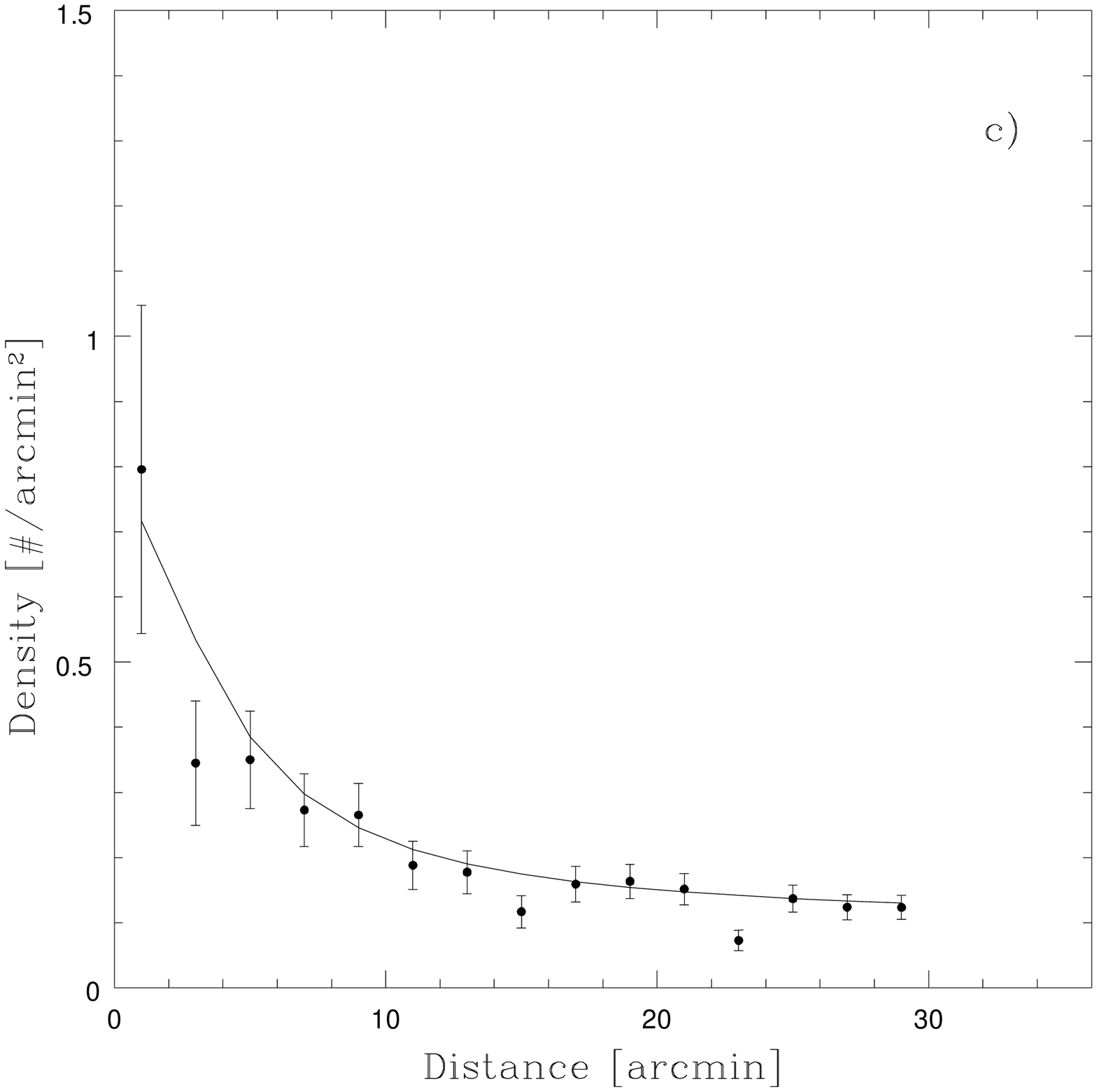}
\caption[]{
c) Galaxy density distribution in A3562, as a function of the distance
from the cluster center.
No background subtraction has been applied in this figure. } 
\label{fig:a3562prof}
\end{figure}
%-------------------------------------------------------------------------------

%-------------------------------------------------------------------------------
\subsection{SC 1329--314}

In Paper I, from the bi--dimensional and redshift distribution of galaxies 
between A3562 and A3558, we presented some evidence that the poor cluster
SC 1329--314 can be considered as formed by two entities. 
In particular, some degree of luminosity segregation might be present 
between the two substructures. In Paper II, on the basis of the ROSAT 
observation of A3558, we confirmed the existence of both potential wells, 
revealing diffuse hot gas emission, as done also by Breen et al. (1994) 
using an Einstein observation.

Breen et al. (1994) dubbed the two groups SC 1329--313 and SC 1327--312 
(named by us SC 1329--314A and SC 1329--314B in Paper I): for simplicity we 
will adopt their terminology.
Although we do not have new redshift information for these systems, the detection
of the diffuse gas can help us to better define the centers and sizes 
of these groups, avoiding (or at least decreasing) the possibility of 
contamination from the overall complex. We decided to use
the positions obtained from the X--ray frame in which the groups are better
centered: therefore for SC 1329--313 the coordinates given by Breen et al. 
[$\alpha(2000)=13^h 31^m 36^s$ and $\delta(2000)=-31^o 48' 46''$] are assumed,
while for SC 1327--312 we prefer the center given in Paper II 
[$\alpha(2000)=13^h 29^m 47^s$ and $\delta(2000)=-31^o 36' 29''$]. For each 
group we selected all galaxies within 10 arcmin from the center. 

The clump SC 1329--313 is more evident as bi--dimensional
overdensity (see Figure 14 of Paper I). The velocity histogram
of the sample obtained with the X--ray center 
shows bimodality, confirmed by the KMM test at $97\%$ confidence level.
The two distributions have \vmed $=14790^{+114}_{-67}$ km/s, 
$\sigma=377^{+93}_{-82}$ km/s and \vmed $=13348^{+69}_{-83}$ km/s, 
$\sigma=276^{+70}_{-61}$ km/s; the number of objects used is
$16$ and $21$, respectively.  Note that the use of the X--ray position
to select galaxies leads to a more reasonable determination of the
velocity dispersion, with respect to the former value of $\sim 900$ km/s
given in Paper I, which evidently was contaminated by galaxies belonging to
the overall structure, hiding the bimodality.

The subcondensation SC 1327--312 appears to be connected to A3558
by a significant bridge of hot gas, signature of strong mareal interactions,
and it is possible that the connection is extending also to A3562.
SC 1327--312, although less evident in the optical isodensity contours,
has an X--ray luminosity that is a factor $\sim 2.5$ greater than that of
SC 1329--313 and only a half of that of A3562 (see Table 2 of Breen et al. 
1994). With $24$ redshifts we obtained \vmed $=14844^{+105}_{-211}$ and $\sigma=
691_{-146}^{+158}$; the X--ray temperature extrapolated from 
eq.(\ref{eq:lubin}) is $3.4$ keV, in agreement with that found by Breen et 
al. (1994).

%-------------------------------------------------------------------------------
\subsection{A3562}

A3562 is a Bautz--Morgan type I cluster, with richness class 2. 
Previous determinations of the mean velocity and velocity dispersion
are \vmed $=14724 \pm 232$ km/s and $\sigma$=$825_{-120}^{+213}$ km/s from 
Quintana et al. (1995), based on $15$ redshifts. David et al. (1993)
reported a hot gas temperature of $KT=3.8\pm 0.5$ keV from an EXOSAT
observation. The spectroscopic sample for this cluster
is rather small and the dynamical situation 
complex: the existence of a ``finger-of-God" effect for the cluster is
not clear and there can be a contamination from a filamentary structure 
extending from $\sim 12000$ to $\sim 14000$ km/s in the East--West direction
(see Figure \ref{fig:wedge}). Moreover, there could be also a contamination 
from SC 1329--313. For these reasons we restricted the sample
to a region with radius smaller than $10$ arcmin from the center
in the analysis of the dynamical parameters: we
find \vmed $=14492^{+225}_{-286}$ km/s and $\sigma=913_{-96}^{+189}$ km/s
with $21$ galaxies, in agreement with the Quintana et al. (1995) results.
In Figure \ref{fig:a3562}a we show the velocity distribution of all galaxies
within one Abell radius from the center of A3562; the 
superimposed Gaussian has the parameters reported above.
 
The velocity dispersion is consistent, within the errors, with the reported 
X--ray temperature: 
however, the small number of redshifts used prevent us from a more detailed 
analysis and we can not exclude a more complicated scenario for this 
cluster, due to contaminations from nearby groups or to the presence of 
substructures.

The parameters of the luminosity function reported in Figure \ref{fig:a3562}b 
are $\alpha=-1.42_{-0.15}^{+0.19}$ and $M^*=-19.84_{-0.61}^{+0.46}$, with an 
effective number of galaxies of 322. 
The $\alpha$ and $M^*$ values are consistent with those of A3558, indicating a 
similar population of galaxies.  

The radial surface density of galaxies has been fitted and the derived 
parameters are $I_o=0.70$ arcmin$^{-2}$ (corresponding to $401$ \hmendue), 
$r_c=3.1$ arcmin ($0.13$ \hmpc), 
$\alpha=0.70$ and $bck=0.10$ arcmin$^{-2}$ ($57$ \hmendue) 
(Figure \ref{fig:a3562prof}c). The fitted background is $\sim 30\%$ higher
than the value obtained from the control area.
The corresponding mass is $M(< 0.75$\hmpc$)= 3.4 \times 10^{14}$ h$^{-1}$ 
M$_{\odot}$ and $M(< 1.5$\hmpc$)= 7.0 \times 10^{14}$ h$^{-1}$ M$_{\odot}$.

%-------------------------------------------------------------------------------
\section{Summary}

In this paper we have presented $174$ new galaxy redshift determinations
in the direction of the core of the Shapley Concentration, a complex 
formed by the three ACO clusters A3558, A3556 and A3562. These velocities
are added to the previous samples of Paper I and of Quintana et al. 
(1995) obtaining a final list of $714$ velocities. The percentage of galaxies
presenting emission lines is $17\%$, consistent with that found in other
clusters.
This new sample confirms the early claim that this structure is formed by
strongly interacting clusters and is elongated (for $\sim 7.5$ \hmpc)
perpendicular to the line of sight. 

With this larger sample, we found that this structure is characterized
by a large number of subcondensations: A3556, A3558 and SC 1329 -313
appear to have subclumps. In particular, we detected a group
with $\sigma=222$ km/s which is probably infalling toward the main 
component of A3556 and which hosts the extended radio source J1324-3138
described in Venturi et al. (1997). The interaction between the media of
the two clumps could be responsible of the ram pressure acting on the 
source. 

Applying the substructure test of Dressler \& Shectman (1988) we found a
significant clump in A3558, in the cluster quadrant with higher X--ray
temperature (Markevitch \& Vikhlinin 1997). Moreover, the two elliptical 
galaxies detected as X--ray sources in the ROSAT map of A3558 (see Paper II) 
are found to reside in a small group southward with respect to the center of 
the cluster, at a distance of $\sim 18$ arcmin. 

For the three ACO clusters of the complex we derived the luminosity 
functions, adopting a new fitting technique which takes into account 
the galaxy density profiles. The dynamical complexity
of this region does not seem to influence noticeably the magnitude
and density distributions of A3558 and A3562, which are similar
to other clusters.
The luminosity function of A3556 has a remarkably fainter
cut off with respect to the ``mean" Schechter function of clusters as estimated 
by Colless (1989) and Lumsden et al. (1997) and appears to have a ``plateau"
for magnitude brighter than $b_J$=15.5: the corresponding galaxies are all
radio emitters. Moreover, from the density profile, galaxies fainter than
$b_J = 18$ appear to have a flatter radial distribution
while the brighter objects are more concentrated toward the cluster center.

The core of the Shapley Concentration, a single connected structure
embedding three cluster cores and other minor subcondensations,
resembles the result of the detailed simulations performed by 
McGlynn \& Fabian (1984) and Roettiger et al. (1993), who studied
cluster--cluster and cluster--group merging, respectively. 

Our general conclusion is that the central part of the Shapley Concentration
is an active region for what concerns dynamical interactions,
where rich clusters are forming through a bottom--up hierachical process. In 
particular, the A3558 cluster complex could be a cluster--cluster collision
seen just after the first core--core encounter, where an intervening 
cluster impacted onto the richer object A3558.
Indeed, the clumpiness found eastward of A3558 could be due to 
the galaxies of this intervening cluster, which is now emerging from the 
main component. However, this major merging event does not seem to have 
modified substantially the galaxy luminosity distribution in the richer
components A3558 and A3562. This implies that, at least during the early stages
of this phenomenon, the galaxy--galaxy interactions are not very important.

This scenario is not unexpected
in a rich environment such as the central part of a supercluster, where 
peculiar velocities of the order of $\sim 1000$ km/s are expected. 
In particular, another similar structure is found $\sim 10$ degrees westward
of the A3558 complex, dominated by the three ACO clusters A3528, A3530 and
A3532. The analysis of a redshift survey toward these objects is in progress.
Moreover, in order to determine the galaxy overdensity of the inner
$10$ \hmpc, we performed a redshift survey of galaxies between clusters:
the results of these analyses will be presented elsewhere (Bardelli et al.,
in preparation).

%-------------------------------------------------------------------------------
%
\section*{Acknowledgements}
We warmly thank Luca Ciotti for helpfull discussions. 
This work has been partially supported through the ASI Contract 95--RS--152.
We thank the referee for useful comments. 
%
%-------------------------------------------------------------------------------

%-------------------------------------------------------------------------------
%

%-------------------------------------------------------------------------------

%-------------------------------------------------------------------------------
% TABLE 2. catalogo redshift

\setcounter{table}{1} 
\begin{table}
\caption[]{ Redshift data }
\begin{flushleft}
\begin{tabular}{lllllll}
\hline\noalign{\smallskip}
\multicolumn{7}{l}{ FIELD 1: $\alpha$ (2000) =~13$^h$~33$^m$~30$^s$ ~
                             $\delta$ (2000) =~-31$^o$~40'~00''        } \\
\noalign{\smallskip}
\hline\noalign{\smallskip}
  \#  & $\alpha$ (2000) & $\delta$ (2000) & $b_J$ & $v$ & err & notes \\
\noalign{\smallskip}
\hline\noalign{\smallskip}
     1 & 13~33~58.02 &   -31~29~24.9 &   14.85 &   10876 & 77 &      \\
     2 & 13~33~49.97 &   -31~43~15.8 &   16.24 &   16162 & 50 &      \\
     3 & 13~34~23.16 &   -31~44~07.5 &   16.33 &   16311 & 50 &      \\
     4 & 13~34~35.87 &   -31~41~07.0 &   16.48 &   11357 & 89 &      \\
     5 & 13~33~50.37 &   -31~28~39.2 &   16.59 &   10994 & 12 & emiss\\
     6 & 13~33~42.03 &   -31~32~37.1 &   16.75 &   15379 & 59 &      \\
     7 & 13~33~37.54 &   -31~33~18.3 &   16.76 &   14289 & 53 &      \\
     8 & 13~33~17.22 &   -31~29~33.6 &   16.88 &   14336 & 64 &      \\
     9 & 13~32~46.44 &   -31~37~03.4 &   16.88 &   12317 &  8 & emiss\\
    10 & 13~33~47.86 &   -31~33~22.2 &   16.97 &   13067 & 44 &      \\
    11 & 13~32~39.21 &   -31~49~52.6 &   17.14 &   12797 & 51 &      \\
    12 & 13~33~11.67 &   -31~40~09.8 &   17.25 &   14269 & 74 &      \\
    13 & 13~32~35.24 &   -31~31~46.2 &   17.28 &   15193 & 18 & emiss\\
    14 & 13~32~43.87 &   -31~47~01.7 &   17.33 &   12836 & 36 & emiss\\
    15 & 13~33~31.93 &   -31~36~40.6 &   17.33 &   15716 & 87 &      \\
    16 & 13~33~37.07 &   -31~38~29.9 &   17.36 &   14995 & 43 &      \\
    17 & 13~34~07.19 &   -31~29~22.0 &   17.39 &   15288 & 77 &      \\
    18 & 13~33~41.03 &   -31~42~52.4 &   17.46 &   13610 & 64 &      \\
    19 & 13~34~38.49 &   -31~40~19.2 &   17.56 &   13154 & 71 &      \\
    20 & 13~34~30.43 &   -31~36~50.2 &   17.59 &   14567 & 83 &      \\
    21 & 13~33~09.49 &   -31~36~12.8 &   17.60 &   14557 & 65 &      \\
    22 & 13~33~13.25 &   -31~34~13.3 &   17.64 &   13382 & 87 &      \\
    23 & 13~32~28.80 &   -31~36~52.4 &   17.67 &   15044 & 87 &      \\
    24 & 13~34~02.55 &   -31~37~00.8 &   17.70 &   14152 &157 &      \\
    25 & 13~33~23.42 &   -31~29~02.1 &   17.78 &   13049 & 53 & emiss\\
    26 & 13~32~37.37 &   -31~44~24.3 &   17.99 &   15444 & 33 & emiss\\
    27 & 13~34~43.36 &   -31~40~23.0 &   18.02 &   10391 &  3 & emiss\\
    28 & 13~33~16.51 &   -31~43~41.7 &   18.13 &   22007 & 28 & emiss\\
    29 & 13~33~59.61 &   -31~47~05.7 &   18.32 &   13110 & 88 &      \\
    30 & 13~33~04.08 &   -31~39~04.6 &   18.32 &   14120 &  7 & emiss\\
    31 & 13~34~01.63 &   -31~26~08.5 &   18.33 &   14683 & 93 &      \\
    32 & 13~34~08.11 &   -31~47~35.5 &   18.34 &   15335 &154 &      \\
    33 & 13~33~08.46 &   -31~29~39.0 &   18.43 &   14256 & 70 &      \\
    34 & 13~32~51.80 &   -31~34~40.9 &   18.45 &   15674 & 52 &      \\
    35 & 13~32~39.43 &   -31~48~36.7 &   18.56 &   14421 & 61 &      \\
    36 & 13~32~22.23 &   -31~44~18.7 &   18.61 &   12340 & 20 & emiss\\
    37 & 13~33~32.91 &   -31~31~18.7 &   18.73 &   15092 &135 &      \\
    38 & 13~33~16.84 &   -31~25~48.3 &   18.79 &   13757 & 31 & emiss\\
    39 & 13~33~22.67 &   -31~33~10.0 &   18.80 &   16122 & 80 &      \\
    40 & 13~32~43.98 &   -31~48~35.6 &   18.83 &   14303 & 82 &      \\
\noalign{\smallskip}
\hline
\end{tabular}
\end{flushleft}
\end{table}
%-------------------------------------------------------------------------------

%-------------------------------------------------------------------------------
\setcounter{table}{1} 
\begin{table}
\caption[]{ cont. }
\begin{flushleft}
\begin{tabular}{lllllll}
\hline\noalign{\smallskip}
\multicolumn{7}{l}{ FIELD 2: $\alpha$ (2000) =~13$^h$~23$^m$~36$^s$ ~
                             $\delta$ (2000) =~-31$^o$~39'~21''        } \\
\noalign{\smallskip}
\hline\noalign{\smallskip}
  \#  & $\alpha$ (2000) & $\delta$ (2000) & $b_J$ & $v$ & err & notes \\
\noalign{\smallskip}
\hline\noalign{\smallskip}
     1 & 13~24~18.66 &   -31~50~33.5 &   17.49 &   16138 & 41 &      \\
     2 & 13~23~43.86 &   -31~47~01.8 &   17.67 &   14325 & 71 &      \\
     3 & 13~22~42.52 &   -31~32~33.2 &   17.82 &   star  &  ~ &      \\
     4 & 13~24~46.36 &   -31~44~46.3 &   17.99 &   15437 & 72 &      \\
     5 & 13~23~49.22 &   -31~42~57.5 &   18.04 &   12998 & 59 &      \\
     6 & 13~24~06.62 &   -31~41~14.0 &   18.17 &   14952 & 50 &      \\
     7 & 13~23~34.48 &   -31~47~19.6 &   18.30 &   14001 & 27 &      \\
     8 & 13~23~41.72 &   -31~28~35.0 &   18.33 &   star  &  ~ &      \\
     9 & 13~22~32.96 &   -31~32~45.7 &   18.52 &   14147 & 95 &      \\
    10 & 13~24~02.96 &   -31~49~45.5 &   18.54 &   13755 & 29 &      \\
    11 & 13~23~06.47 &   -31~37~07.3 &   18.62 &   star  &  ~ &      \\
    12 & 13~24~31.96 &   -31~41~15.7 &   18.66 &   14117 & 37 &      \\
    13 & 13~24~04.91 &   -31~47~47.5 &   18.67 &   15125 & 34 &      \\
    14 & 13~24~40.30 &   -31~38~42.2 &   18.71 &   14296 & 35 &      \\
    15 & 13~23~35.43 &   -31~52~01.2 &   18.72 &   14900 & 34 &      \\
    16 & 13~23~17.04 &   -31~46~16.5 &   18.79 &   star  &  ~ &      \\
    17 & 13~24~10.75 &   -31~40~03.8 &   18.81 &   star  &  ~ &      \\
    18 & 13~23~04.76 &   -31~31~43.9 &   18.82 &   27265 & 95 &      \\
    19 & 13~24~36.28 &   -31~47~41.1 &   18.83 &   35322 & 64 &      \\
    20 & 13~23~48.15 &   -31~49~54.0 &   18.86 &   15074 & 27 &      \\
    21 & 13~22~39.07 &   -31~48~01.5 &   18.87 &   star  &  ~ &      \\
    22 & 13~24~11.26 &   -31~30~57.6 &   18.89 &   13387 & 41 &      \\
    23 & 13~23~10.74 &   -31~50~11.5 &   18.89 &   14019 & 31 &      \\
    24 & 13~23~54.55 &   -31~37~21.1 &   18.90 &   55144 & 76 &      \\
    25 & 13~23~50.14 &   -31~35~19.9 &   18.91 &   13986 & 28 &      \\
    26 & 13~24~43.04 &   -31~32~39.4 &   18.93 &   13919 & 32 &      \\
    27 & 13~22~36.07 &   -31~34~13.4 &   18.94 &   star  &  ~ &      \\
    28 & 13~23~44.51 &   -31~30~31.7 &   18.97 &   12864 & 64 &      \\
    29 & 13~22~40.55 &   -31~40~24.5 &   18.98 &   star  &  ~ &      \\
    30 & 13~23~38.16 &   -31~40~52.5 &   18.99 &   star  &  ~ &      \\
    31 & 13~22~46.65 &   -31~44~03.1 &   19.00 &   14381 & 65 &      \\
    32 & 13~24~17.69 &   -31~29~21.9 &   19.00 &   star  &  ~ &      \\
    33 & 13~23~40.96 &   -31~41~56.5 &   19.02 &   16443 & 39 & emiss\\
    34 & 13~24~37.17 &   -31~44~21.3 &   19.03 &   13967 & 85 &      \\
    35 & 13~23~35.83 &   -31~53~29.0 &   19.04 &   star  &  ~ &      \\
    36 & 13~22~56.94 &   -31~38~54.8 &   19.04 &   star  &  ~ &      \\
    37 & 13~23~25.43 &   -31~32~48.1 &   19.06 &   40320 & 43 &      \\
    38 & 13~24~25.97 &   -31~42~16.1 &   19.08 &   star  &  ~ &      \\
    39 & 13~23~19.77 &   -31~24~18.0 &   19.08 &   16480 &  4 & emiss\\
    40 & 13~23~07.65 &   -31~39~15.1 &   19.09 &   49858 & 79 &      \\
    41 & 13~24~24.95 &   -31~49~49.4 &   19.11 &   star  &  ~ &      \\
    42 & 13~23~30.73 &   -31~34~38.7 &   19.12 &   star  &  ~ &      \\
    43 & 13~24~42.09 &   -31~32~01.6 &   19.13 &   14339 & 38 &      \\
\noalign{\smallskip}
\hline
\end{tabular}
\end{flushleft}
\end{table}
%-------------------------------------------------------------------------------

%-------------------------------------------------------------------------------
\setcounter{table}{1} 
\begin{table}
\caption[]{ cont. }
\begin{flushleft}
\begin{tabular}{lllllll}
\hline\noalign{\smallskip}
\multicolumn{7}{l}{ FIELD 3: $\alpha$ (2000) =~13$^h$~29$^m$~45$^s$ ~
                             $\delta$ (2000) =~-32$^o$~07'~43''        } \\
\noalign{\smallskip}
\hline\noalign{\smallskip}
  \#  & $\alpha$ (2000) & $\delta$ (2000) & $b_J$ & $v$ & err & notes \\
\noalign{\smallskip}
\hline\noalign{\smallskip}
     1 & 13~30~03.99 &   -32~02~10.2 &   15.57 &    4537 & 20 & emiss\\
     2 & 13~29~55.95 &   -32~08~31.4 &   15.72 &    3873 & 63 &      \\
     3 & 13~29~16.93 &   -32~02~35.0 &   16.15 &   13834 & 29 &      \\
     4 & 13~30~34.64 &   -32~15~52.3 &   16.22 &    4027 & 33 &      \\
     5 & 13~30~20.46 &   -32~05~50.1 &   16.26 &   12770 & 34 &      \\
     6 & 13~30~52.08 &   -32~08~56.3 &   16.55 &   12932 & 31 &      \\
     7 & 13~28~50.84 &   -32~07~54.3 &   16.86 &   15073 & 57 &      \\
     8 & 13~30~21.19 &   -31~55~44.0 &   16.92 &   14204 & 28 &      \\
     9 & 13~29~10.37 &   -32~05~24.3 &   17.10 &   13816 & 49 &      \\
    10 & 13~30~36.42 &   -32~18~37.8 &   17.12 &   13040 & 26 &      \\
    11 & 13~29~54.10 &   -32~01~00.5 &   17.25 &   14012 & 23 &      \\
    12 & 13~30~06.42 &   -31~56~26.9 &   17.31 &   star  &  ~ &      \\
    13 & 13~30~18.90 &   -32~01~54.5 &   17.33 &   12927 & 39 &      \\
    14 & 13~30~40.67 &   -32~05~35.6 &   17.36 &   12961 & 47 &      \\
    15 & 13~29~01.36 &   -32~01~41.7 &   17.43 &   14047 & 52 &      \\
    16 & 13~30~47.52 &   -32~02~36.7 &   17.44 &   13435 & 28 &      \\
    17 & 13~30~07.95 &   -32~22~41.4 &   17.44 &   14618 & 31 &      \\
    18 & 13~30~38.67 &   -32~03~28.9 &   17.51 &   13100 & 27 &      \\
    19 & 13~28~51.47 &   -32~04~18.2 &   17.65 &   13479 & 28 &      \\
    20 & 13~29~58.21 &   -32~12~27.1 &   17.68 &   14060 & 28 &      \\
    21 & 13~28~41.84 &   -32~16~26.9 &   17.72 &   13779 & 45 &      \\
    22 & 13~30~12.67 &   -32~07~24.2 &   17.73 &   14027 & 25 &      \\
    23 & 13~29~33.66 &   -31~53~26.6 &   17.79 &   14525 & 14 & emiss\\
    24 & 13~28~58.74 &   -32~03~52.8 &   17.82 &   14467 & 20 &      \\
    25 & 13~29~21.63 &   -31~59~35.7 &   17.85 &   star  &  ~ &      \\
    26 & 13~30~01.71 &   -31~56~49.4 &   17.99 &   14122 & 32 &      \\
    27 & 13~28~46.44 &   -32~10~01.5 &   18.03 &   13710 & 11 & emiss\\
    28 & 13~30~02.02 &   -32~15~19.7 &   18.06 &   14936 & 36 &      \\
    29 & 13~30~15.46 &   -32~03~59.7 &   18.07 &   13835 & 31 &      \\
    30 & 13~30~02.85 &   -31~56~25.4 &   18.13 &   14855 & 37 &      \\
    31 & 13~29~36.89 &   -32~20~47.9 &   18.34 &   star  &  ~ &      \\
    32 & 13~30~00.67 &   -32~18~33.1 &   18.46 &   12288 &  9 & emiss\\
    33 & 13~29~16.02 &   -32~16~37.7 &   18.49 &   38646 & 47 &      \\
    34 & 13~29~49.34 &   -32~22~16.8 &   18.49 &   52895 & 49 &      \\
    35 & 13~30~51.61 &   -32~11~43.6 &   18.54 &   12958 & 81 &      \\
    36 & 13~29~06.33 &   -31~56~37.5 &   18.60 &   12705 &  8 & emiss\\
    37 & 13~30~29.12 &   -32~00~47.8 &   18.70 &   12884 & 68 &      \\
    38 & 13~28~54.26 &   -31~56~25.3 &   18.70 &   15473 & 33 &      \\
    39 & 13~29~37.35 &   -31~55~57.9 &   18.72 &   14238 & 81 &      \\
    40 & 13~30~00.49 &   -31~52~28.6 &   18.73 &   13430 & 18 & emiss\\
    41 & 13~30~29.40 &   -32~13~53.2 &   18.80 &    4043 &185 &      \\
    42 & 13~28~53.62 &   -32~17~27.7 &   18.81 &   43458 & 28 &      \\
\noalign{\smallskip}
\hline
\end{tabular}
\end{flushleft}
\end{table}
%-------------------------------------------------------------------------------

%-------------------------------------------------------------------------------
\setcounter{table}{1} 
\begin{table}
\caption[]{ cont. }
\begin{flushleft}
\begin{tabular}{lllllll}
\hline\noalign{\smallskip}
\multicolumn{7}{l}{ FIELD 4: $\alpha$ (2000) =~13$^h$~27$^m$~15$^s$ ~
                             $\delta$ (2000) =~-32$^o$~05'~00''        } \\
\noalign{\smallskip}
\hline\noalign{\smallskip}
  \#  & $\alpha$ (2000) & $\delta$ (2000) & $b_J$ & $v$ & err & notes \\
\noalign{\smallskip}
\hline\noalign{\smallskip}
     1 & 13~26~55.63 &   -31~58~13.9 &   15.68 &    8058 & 74 &      \\
     2 & 13~26~48.59 &   -31~59~47.1 &   15.75 &   14779 & 46 &      \\
     3 & 13~26~52.72 &   -32~00~31.3 &   16.19 &   14004 & 40 &      \\
     4 & 13~27~24.31 &   -32~18~01.4 &   16.34 &   14329 & 78 &      \\
     5 & 13~27~06.52 &   -31~51~40.4 &   16.38 &   15407 & 33 &      \\
     6 & 13~26~57.42 &   -32~11~53.8 &   16.57 &   13982 & 86 &      \\
     7 & 13~27~15.64 &   -32~17~10.0 &   16.61 &   13057 & 58 &      \\
     8 & 13~28~08.01 &   -32~08~48.4 &   16.69 &   21550 & 82 &      \\
     9 & 13~27~12.92 &   -31~55~54.0 &   16.73 &   14623 & 41 &      \\
    10 & 13~26~57.53 &   -31~56~07.6 &   17.25 &   14662 & 34 &      \\
    11 & 13~27~03.20 &   -31~52~23.9 &   17.52 &   14947 & 34 &      \\
    12 & 13~26~33.73 &   -31~56~56.1 &   17.52 &   15217 &101 &      \\
    13 & 13~27~51.39 &   -31~57~00.5 &   17.59 &   15152 & 33 &      \\
    14 & 13~27~19.84 &   -31~54~50.8 &   17.66 &   13619 & 43 &      \\
    15 & 13~27~18.62 &   -31~51~09.5 &   17.90 &   15568 & 54 &      \\
    16 & 13~26~53.38 &   -31~52~26.1 &   17.94 &   14547 & 30 &      \\
    17 & 13~26~26.65 &   -31~55~32.9 &   17.98 &   14911 & 81 &      \\
    18 & 13~27~14.00 &   -31~59~15.5 &   18.02 &   14575 & 62 &      \\
    19 & 13~26~58.45 &   -32~14~54.4 &   18.17 &   16589 & 51 &      \\
    20 & 13~27~07.35 &   -32~00~12.6 &   18.19 &   15904 & 64 &      \\
    21 & 13~27~36.20 &   -32~03~03.5 &   18.39 &   star  &  ~ &      \\
    22 & 13~26~59.92 &   -31~57~54.8 &   18.41 &   12843 &104 &      \\
    23 & 13~28~10.82 &   -32~12~52.6 &   18.44 &   43169 & 33 &      \\
    24 & 13~27~49.44 &   -31~59~40.8 &   18.49 &   13553 & 27 &      \\
    25 & 13~27~03.48 &   -32~16~26.4 &   18.49 &   16425 & 30 &      \\
    26 & 13~27~09.48 &   -32~09~18.7 &   18.52 &   13791 & 88 &      \\
    27 & 13~26~49.74 &   -31~51~24.8 &   18.58 &   star  &  ~ &      \\
    28 & 13~28~23.33 &   -32~06~55.4 &   18.66 &   21934 & 72 &      \\
    29 & 13~27~10.94 &   -31~54~13.8 &   18.67 &   13140 & 50 &      \\
    30 & 13~26~54.83 &   -31~59~49.3 &   18.67 &   14569 & 77 &      \\
    31 & 13~27~14.63 &   -31~53~43.0 &   18.68 &    7959 & 10 & emiss\\
    32 & 13~28~16.84 &   -32~12~09.5 &   18.68 &   13134 & 36 &      \\
    33 & 13~28~14.71 &   -32~08~30.1 &   18.70 &   star  &  ~ &      \\
    34 & 13~28~14.14 &   -32~12~33.2 &   18.72 &   star  &  ~ &      \\
    35 & 13~28~07.50 &   -32~14~43.7 &   18.75 &   43613 & 43 &      \\
    36 & 13~27~32.96 &   -32~11~36.4 &   18.81 &   14474 & 94 &      \\
    37 & 13~27~19.38 &   -31~54~14.3 &   18.81 &   37199 & 15 & emiss\\
    38 & 13~28~01.22 &   -32~04~34.6 &   18.85 &   14454 & 48 &      \\
    39 & 13~26~45.57 &   -32~14~19.0 &   18.85 &   star  &  ~ &      \\
    40 & 13~26~17.24 &   -32~09~22.3 &   18.92 &   14344 & 59 &      \\
    41 & 13~26~27.61 &   -31~53~43.8 &   18.97 &   58296 & 32 &      \\
    42 & 13~26~48.09 &   -32~18~21.8 &   18.98 &   43351 & 42 &      \\
    43 & 13~27~45.88 &   -31~51~17.5 &   19.00 &   14562 &  9 & emiss\\
    44 & 13~27~20.39 &   -32~12~15.8 &   19.02 &   13359 & 56 &      \\
\noalign{\smallskip}
\hline
\end{tabular}
\end{flushleft}
\end{table}
%-------------------------------------------------------------------------------

%-------------------------------------------------------------------------------
\setcounter{table}{1} 
\begin{table}
\caption[]{ cont. }
\begin{flushleft}
\begin{tabular}{lllllll}
\hline\noalign{\smallskip}
\multicolumn{7}{l}{ FIELD 5: $\alpha$ (2000) =~13$^h$~32$^m$~15$^s$ ~
                             $\delta$ (2000) =~-32$^o$~12'~17''        } \\
\noalign{\smallskip}
\hline\noalign{\smallskip}
  \#  & $\alpha$ (2000) & $\delta$ (2000) & $b_J$ & $v$ & err & notes \\
\noalign{\smallskip}
\hline\noalign{\smallskip}
     1 & 13~31~19.38 &   -31~13~45.0 &   16.35 &   15710 & 27 &      \\
     2 & 13~31~06.07 &   -31~16~57.7 &   16.92 &    3958 & 10 & emiss\\
     3 & 13~31~14.90 &   -31~11~53.8 &   16.99 &   15607 & 35 &      \\
     4 & 13~31~13.73 &   -31~13~02.4 &   17.01 &   14239 & 28 &      \\
     5 & 13~32~09.09 &   -31~03~27.0 &   17.07 &   15729 & 38 &      \\
     6 & 13~32~28.46 &   -31~03~50.1 &   17.17 &   13852 & 31 &      \\
     7 & 13~31~47.71 &   -31~25~39.6 &   17.20 &   14214 & 32 &      \\
     8 & 13~33~00.93 &   -31~13~29.6 &   17.53 &   11028 & 10 & emiss\\
     9 & 13~31~13.75 &   -31~16~28.3 &   17.68 &   13105 & 26 &      \\
    10 & 13~31~22.32 &   -31~22~10.5 &   17.70 &   14027 & 13 & emiss\\
    11 & 13~32~11.64 &   -31~02~21.2 &   17.70 &   13425 & 30 &      \\
    12 & 13~32~30.00 &   -31~03~16.6 &   17.74 &   13772 & 48 &      \\
    13 & 13~33~20.34 &   -31~15~17.0 &   17.78 &   50004 & 54 &      \\
    14 & 13~31~55.90 &   -31~14~19.6 &   17.82 &   13717 & 29 &      \\
    15 & 13~32~55.16 &   -31~18~13.8 &   17.87 &   star  &  ~ &      \\
    16 & 13~32~05.32 &   -31~12~29.4 &   17.90 &   star  &  ~ &      \\
    17 & 13~32~31.63 &   -31~07~27.1 &   18.02 &   star  &  ~ &      \\
    18 & 13~31~42.47 &   -31~17~57.7 &   18.09 &   14056 & 11 & emiss\\
    19 & 13~31~55.52 &   -31~01~43.2 &   18.24 &   15235 & 29 &      \\
    20 & 13~32~12.91 &   -31~12~44.7 &   18.30 &   star  &  ~ &      \\
    21 & 13~32~38.22 &   -31~01~41.8 &   18.32 &   13189 &  4 & emiss\\
    22 & 13~31~35.61 &   -31~23~28.2 &   18.32 &   star  &  ~ &      \\
    23 & 13~31~05.56 &   -31~12~57.7 &   18.33 &   star  &  ~ &      \\
    24 & 13~32~33.41 &   -31~20~51.0 &   18.33 &   star  &  ~ &      \\
    25 & 13~32~09.41 &   -31~09~24.3 &   18.35 &   star  &  ~ &      \\
    26 & 13~33~21.74 &   -31~05~12.2 &   18.36 &   33108 & 33 &      \\
    27 & 13~31~19.20 &   -31~01~43.7 &   18.45 &   39256 & 90 &      \\
    28 & 13~32~29.30 &   -31~06~18.7 &   18.48 &   30796 & 58 &      \\
    29 & 13~32~36.01 &   -31~08~39.4 &   18.48 &   star  &  ~ &      \\
    30 & 13~32~28.11 &   -31~25~37.7 &   18.49 &   27119 &108 &      \\
    31 & 13~33~27.82 &   -31~12~54.2 &   18.53 &   13059 & 13 & emiss\\
    32 & 13~31~42.01 &   -31~10~43.3 &   18.57 &   star  &  ~ &      \\
    33 & 13~32~20.37 &   -31~12~01.3 &   18.63 &   star  &  ~ &      \\
    34 & 13~32~30.78 &   -31~02~07.8 &   18.66 &   13334 & 11 & emiss\\
    35 & 13~32~43.80 &   -31~23~09.0 &   18.66 &   star  &  ~ &      \\
    36 & 13~32~25.83 &   -31~20~17.1 &   18.71 &   14919 & 39 &      \\
    37 & 13~32~52.33 &   -31~20~31.3 &   18.76 &   16546 & 35 &      \\
    38 & 13~33~12.04 &   -31~08~19.1 &   18.80 &   star  &  ~ &      \\
    39 & 13~31~01.76 &   -31~10~49.1 &   18.80 &   star  &  ~ &      \\
    40 & 13~32~15.09 &   -31~13~29.8 &   18.81 &   44703 & 74 &      \\
    41 & 13~31~54.07 &   -31~01~04.6 &   18.81 &   star  &  ~ &      \\
    42 & 13~31~53.30 &   -31~21~46.1 &   18.88 &   star  &  ~ &      \\
    43 & 13~31~22.88 &   -31~06~19.0 &   18.90 &   14483 & 53 &      \\
    44 & 13~33~09.95 &   -31~22~42.2 &   18.93 &   star  &  ~ &      \\
    45 & 13~31~56.36 &   -31~08~53.2 &   18.94 &   15316 &  5 & emiss\\
\noalign{\smallskip}
\hline
\end{tabular}
\end{flushleft}
\end{table}
%-------------------------------------------------------------------------------


\begin{thebibliography}{99}

\bibitem{} Abell, G.O., Corwin Jr., H.G., Olowin, R.P., 1989, 
           ApJSS 70, 1 [ACO catalogue]

\bibitem{} Allen, D.A., Norris, R.P., Staveley--Smith, L., Meadows, V.S., 
           Roche, P.F., 90, Nat 343, 45.

\bibitem{} Ashman, K.M., Bird, C.M., Zepf, S.E., 1994, AJ 108, 2348

\bibitem{} Bahcall, J.N., Tremaine, S., 1981, ApJ 244, 805 

\bibitem{} Bardelli, S., Zucca, E., Vettolani, G., Zamorani, G., 
           Scaramella, R., Collins, C.A., MacGillivray, H.T., 1994,
           MNRAS 267, 665 [Paper I]

\bibitem{} Bardelli, S., Zucca, E., Malizia, A., Zamorani, G., Scaramella, R.,
           Vettolani, G., 1996, A\&A 305, 435 [Paper II]

\bibitem{} Bardelli, S., Zucca, E., Vettolani, G., Zamorani, G., Scaramella,
           R., 1997, Proceedings of Observational Cosmology: from 
           Galaxies to Galaxy Systems, Astroph. Letters and Comm., in press

\bibitem{} Beers, T.C., Flynn, K., Gebhardt, K., 1990, AJ 100, 32

\bibitem{} Binney, J., Tremaine, S., 1987, Galactic Dynamics, Princeton
           University press, Princeton, p.205

\bibitem{} Bird, C.M., Beers, T.C., 1993, AJ 105, 1596

\bibitem{} Bird, C.M., 1994, AJ 107, 1637

\bibitem{} Biviano, A., Girardi, M., Giuricin, G., Mardirossian, F.,
           Mezzetti, M., 1992, ApJ 396, 35

\bibitem{} Biviano, A., Girardi, M., Giuricin, G., Mardirossian, F.,
           Mezzetti, M., 1993, ApJ 411, L13

\bibitem{} Biviano, A., Katgert, P., Mazure, A., Moles, M., den Hartog, R.,
    Focardi, P., 1997a, Proceedings of Observational Cosmology: from 
    Galaxies to Galaxy Systems, Astroph. Letters and Comm., in press

\bibitem{} Biviano, A., Katgert, P., Mazure, A., Moles, M., den Hartog, R.,
           Perea, J., Focardi, P., 1997b, A\&A 321, 84

\bibitem{} Branchini, E., Plionis, M., 1996, ApJ 460, 569

\bibitem{} Breen, J., Raychaudhury, S., Forman, W., Jones, C., 1994, ApJ
           424, 59

\bibitem{} Briel, U.G., Henry, J.P., Schwarz, R.A., B\"ohringer, H., Ebeling,
     H., Edge, A.C., Hartner, G.D., Schindler, S., Tr\"umper, J., Voges, W.,
     1991, A\&A 246, L10

\bibitem{} Burns, J.O., Roettiger, K., Ledlow, M., Klypin, A., 1994, 
           ApJ 427, L90

\bibitem{} Burstein, D., Heiles, C., 1984, ApJSS 54, 33

\bibitem{} Colless, M., 1989, MNRAS 237, 799

\bibitem{} Danese, L., De Zotti, G., di Tullio, G., 1980, A\&A 82, 322

\bibitem{} Dantas, C.C., de Carvalho, R.R., Capelato, H.V., Mazure, A.,
           1997, ApJ 485, 447

\bibitem{} David, L.P., Slyz, A., Jones, C., Forman, W., Vrtilek, S.D., 
           Arnaud, K.A., 1993, ApJ 412, 479

\bibitem{} Dressler, A., Shectman, S.A., 1988, AJ 95, 985

\bibitem{} Feretti, L., B\"ohringer, H., Giovannini, G., Neumann, D., 1997,
           A\&A 317, 432

\bibitem{} Feretti, L., Giovannini, G., 1996, in Extragalactic Radio Sources,
           IAU Symposium 175, Ekers, R., Fanti, C., Padrielli, L. eds.,
           Kluwer academic publ., Dordrecht, p.334 

\bibitem{} Girardi, M., Biviano, A., Giuricin, G., Mardirossian, F., Mezzetti,
           M., 1995, ApJ 438, 527

\bibitem{} Girardi, M., Escalera, E., Fadda, D., Giuricin, G., Mardirossian, F.,
           Mezzetti, M., 1997, ApJ 482, 41

\bibitem{} Heydon--Dumbleton, N.H., Collins, C.A., MacGillivray, H.T., 1989,
           MNRAS 238, 379

\bibitem{} Kurtz, M.J., Mink, D.J., Wyatt, W.F., Fabricant, D.G., Torres, G., 
     Kriss, G.A., Tonry, J.L., 1992, in Astronomical Data Analysis
     Software and Systems I, Worrall, D.M., Biemesderfer, C., and Barnes, J.,
     eds., ASP conference series vol.25, p.432. 

\bibitem{} Ikebe, Y., Makishima, K., Ezawa, H., et al., 1997, ApJ 481, 660 

\bibitem{} Lahav, O., Edge, A.C., Fabian, A.C., Putney, A., 1989, MNRAS 238, 881

\bibitem{} Lubin, L.M., Bahcall, N.A., 1993, ApJ 415, L17

\bibitem{} Lumsden, S.L., Collins, C.A., Nichol, R.C., Eke, V.R.,
           Guzzo, L., 1997, MNRAS in press (preprint astro-ph/9705120) 
 
\bibitem{} Lund, G., 1986, OPTOPUS - ESO operating manual N.6

\bibitem{} Malumuth, E.M., Kriss, G.A., Van Dyke Dixon, W., Ferguson, H.C.,
           Ritchie, C., 1992, AJ 104, 495 

\bibitem{} Markevitch, M., Vikhlinin, A., 1997, ApJ 474, 84

\bibitem{} McGlynn, T.A., Fabian, A.C., 1984, MNRAS 208, 709

\bibitem{} Metcalfe, N., Godwin, J.G., Spenser, S.D., 1987, MNRAS 225, 581

\bibitem{} Metcalfe, N., Godwin, J.G., Peach, J.V., 1994, MNRAS 267, 431

\bibitem{} Quintana, H., Ramirez, A., Melnick, J., Raychaudhury, S., Szelak,
           E., 1995, AJ 110, 463

\bibitem{} Raychaudhury, S., Fabian, A.C., Edge, A.C., Jones, C., Forman, W.,
           1991, MNRAS 248, 101

\bibitem{} Roettiger, K., Burns, J.O., Loken, C., 1993, ApJ 407, L53

\bibitem{} Scaramella, R., Baiesi--Pillastrini, G., Chincarini, G., 
           Vettolani, G., Zamorani, G., 1989, Nature 338, 562

\bibitem{} Schechter, P., 1976, ApJ 203, 297

\bibitem{} Shapley, H., 1930, Bull. Harvard Obs. 874, 9

\bibitem{} Stein, P., 1996, A\&ASS 116, 203

\bibitem{} Stein, P., 1997, A\&A 317, 670

\bibitem{} Teague, P.F., Carter, D., Gray, P.M., 1990, ApJSS 72, 715

\bibitem{} The L.S., White, S.D.M., 1986, AJ 92, 1248

\bibitem{} Tremaine, S., in Dynamics and Interactions of Galaxies, 1990,
           Wielen, R., ed., Springer Verlag, Berlin, p.394

\bibitem{} Venturi, T., Bardelli, S., Morganti, R., Hunstead, R.W., 1997,
           MNRAS  285, 898

\bibitem{} West, M.J., Oemler, A., Dekel, A., 1988, ApJ 327, 1
         
\bibitem{} Wyse, R.F., Gilmore G., 1992, MNRAS 257, 1

\bibitem{} Yentis, D.J., et al., 1992, in Digitized Optical Sky Surveys,
         MacGillivray, H.T., Collins, C.A., eds., Kluwer, Dodrecht, p.67

\bibitem{} Zucca, E., Zamorani, G., Scaramella, R., Vettolani, G., 1993, 
           ApJ 407, 470

\bibitem{} Zucca, E., Pozzetti, L., Zamorani, G., 1994 MNRAS 269, 953

\end{thebibliography}
\end{document}